\documentclass[fleqn,usenatbib]{mnras}

\usepackage{newtxtext,newtxmath}

\usepackage[T1]{fontenc}
\usepackage{ae,aecompl}

\usepackage{graphicx}	
\usepackage{amsmath}	
\usepackage{amssymb}	
\usepackage{xcolor}

\usepackage{comment}

\usepackage{ulem}

\usepackage{caption}
\usepackage{longtable}

\title[\textit{El Diablo}: A double-horned TDE]{The Tidal Disruption Event AT\,2018hyz I: Double-peaked emission lines and a flat Balmer decrement}

\author[P. Short et al]{\href{https://orcid.org/0000-0002-5096-9464}{P. Short$^{1}$}\thanks{Contact e-mail: \href{mailto:pshort@roe.ac.uk}{pshort@roe.ac.uk}}, 
\href{https://orcid.org/0000-0002-2555-3192}{M. Nicholl$^{2,1}$}, 
\href{https://orcid.org/0000–0002–3134–6093}{A. Lawrence$^{1}$}, 
\href{https://orcid.org/0000-0001-6395-6702}{S. Gomez$^{3}$},
\href{https://orcid.org/0000-0001-7090-4898}{I. Arcavi$^{4,5}$}, 
\href{https://orcid.org/0000-0002-4043-9400}{T. Wevers$^{6}$}, 
\newauthor 
\href{https://orcid.org/0000-0002-8597-0756}{G. Leloudas$^{7}$},
\href{https://orcid.org/0000-0001-6797-1889}{S. Schulze$^{8}$},
\href{https://orcid.org/0000-0003-0227-3451}{J. P. Anderson$^{9}$}, 
\href{https://orcid.org/0000-0002-9392-9681}{E. Berger$^{3}$}, 
\href{https://orcid.org/0000-0003-0526-2248}{P.K. Blanchard$^{10}$}, 
J. Burke$^{11,12}$,
\newauthor 
\href{https://orcid.org/0000-0002-5870-0443}{N. Castro Segura,$^{13}$}
\href{https://orcid.org/0000-0002-0326-6715}{P. Charalampopoulos$^{7}$}, 
\href{https://orcid.org/0000-0002-7706-5668}{R. Chornock$^{14}$}, 
\href{https://orcid.org/0000-0002-1296-6887}{L. Galbany$^{15}$}, 
\href{https://orcid.org/0000-0002-1650-1518}{M. Gromadzki$^{16}$},
\newauthor
\href{https://orcid.org/0000-0002-3680-9712}{L. J. Herzog$^{14}$},
\href{https://orcid.org/0000-0002-1125-9187}{D. Hiramatsu$^{11,12}$}, 
\href{https://orcid.org/0000-0003-1728-0304}{Keith Horne$^{17}$}, 
\href{https://orcid.org/0000-0002-0832-2974}{G. Hosseinzadeh$^{3}$}, 
\href{https://orcid.org/0000-0003-4253-656X}{D. Andrew Howell$^{11,12}$},
\newauthor
N. Ihanec$^{15}$, 
\href{https://orcid.org/0000-0002-3968-4409}{C. Inserra$^{18}$}, 
\href{https://orcid.org/0000-0001-8257-3512}{E. Kankare$^{19}$}, 
\href{https://orcid.org/0000-0002-9770-3508}{K. Maguire$^{20}$}, 
\href{https://orcid.org/0000-0001-5807-7893}{C. McCully$^{12}$},
\newauthor
\href{https://orcid.org/0000-0003-3939-7167}{T. E. M{\"u}ller Bravo$^{13}$},
\href{https://orcid.org/0000-0001-6286-1744}{F. Onori$^{21}$}, 
\href{https://orcid.org/0000-0003-1546-6615}{J. Sollerman$^{22}$}, 
D. R. Young$^{23}$
\\
$^{1}$Institute for Astronomy, University of Edinburgh, Royal Observatory, Blackford Hill, Edinburgh EH9 3HJ, UK\\
$^{2}$Birmingham Institute for Gravitational Wave Astronomy and School of Physics and Astronomy, University of Birmingham, Birmingham B15 2TT, UK\\
$^{3}$Center for Astrophysics | Harvard \& Smithsonian, 60 Garden Street, Cambridge, MA 02138-1516, USA\\
$^{4}$The School of Physics and Astronomy, Tel Aviv University, Tel Aviv 69978, Israel\\
$^{5}$CIFAR Azrieli Global Scholars program, CIFAR, Toronto, Canada\\
$^{6}$Institute of Astronomy, University of Cambridge, Madingley Road, Cambridge CB3 0HA, UK\\
$^{7}$DTU Space, National Space Institute, Technical University of Denmark, Elektrovej 327, DK-2800 Kgs. Lyngby, Denmark\\
$^{8}$Department of Particle Physics and Astrophysics, Weizmann Institute of Science, 234 Herzl St, 76100 Rehovot, Israel\\
$^{9}$European Southern Observatory, Alonso de C\'ordova 3107, Casilla 19, Santiago, Chile\\
$^{10}$Center for Interdisciplinary Exploration and Research in Astrophysics and Department of Physics and Astronomy, \\Northwestern University, 2145 Sheridan Road, Evanston, IL 60208-3112, USA\\
$^{11}$Department of Physics, University of California, Santa Barbara, CA 93106-9530, USA\\
$^{12}$Las Cumbres Observatory, 6740 Cortona Dr, Suite 102, Goleta, CA 93117-5575, USA\\
$^{13}$School of Physics and Astronomy, University of Southampton, Southampton, SO17 1BJ, UK\\
$^{14}$Astrophysical Institute, Department of Physics and Astronomy, 251B Clippinger Lab, Ohio University, Athens, OH 45701-2942, USA\\
$^{15}$Departamento de F\'isica Te\'orica y del Cosmos, Universidad de Granada, E-18071 Granada, Spain.\\
$^{16}$Astronomical Observatory, University of Warsaw, Al. Ujazdowskie 4, 00-478 Warszawa, Poland\\
$^{17}$University of St~Andrews, SUPA School of Physics \& Astronomy, North Haugh, St~Andrews, KY16~9SS, Scotland, UK.\\
$^{18}$School of Physics \& Astronomy, Cardiff University, Queens Buildings, The Parade, Cardiff, CF24 3AA, UK\\
$^{19}$Department of Physics and Astronomy, University of Turku, FI-20014 Turku, Finland\\
$^{20}$School of Physics, Trinity College Dublin, College Green, Dublin 2, Ireland.\\
$^{21}$Istituto di Astrofisica e Planetologia Spaziali (INAF), via del Fosso del Cavaliere 100, Roma I-00133, Italy\\
$^{22}$The Oskar Klein Centre, Department of Astronomy, AlbaNova, SE-106 91 Stockholm , Sweden\\
$^{23}$Astrophysics Research Centre, School of Mathematics and Physics, Queen's University Belfast, Belfast BT7 1NN, UK
}

\date{Accepted XXX. Received YYY; in original form ZZZ}

\pubyear{2020}

\begin{document}
\label{firstpage}
\pagerange{\pageref{firstpage}--\pageref{lastpage}}
\maketitle

\begin{abstract}
We present results from spectroscopic observations of AT\,2018hyz, a transient discovered by the ASAS-SN survey at an absolute magnitude of $M_V\sim -20.2$ mag, in the nucleus of a quiescent galaxy with strong Balmer absorption lines. AT\,2018hyz shows a blue spectral continuum and broad emission lines, consistent with previous TDE candidates. High cadence follow-up spectra show broad Balmer lines and He I in early spectra, with He II making an appearance after $\sim70-100$ days. The Balmer lines evolve from a smooth broad profile, through a boxy, asymmetric double-peaked phase consistent with accretion disc emission, and back to smooth at late times. The Balmer lines are unlike typical AGN in that they show a flat Balmer decrement (H$\alpha$/H$\beta\sim1.5$), suggesting the lines are collisionally excited rather than being produced via photo-ionisation. The flat Balmer decrement together with the complex profiles suggest that the emission lines originate in a disc chromosphere, analogous to those seen in cataclysmic variables. The low optical depth of material due to a possible partial disruption may be what allows us to observe these double-peaked, collisionally excited lines. The late appearance of He II may be due to an expanding photosphere or outflow, or late-time shocks in debris collisions.
\end{abstract}




\section{Introduction}
A tidal disruption event (TDE) is predicted to occur when a star passes within the tidal radius of a supermassive black hole (SMBH). In this region the tidal force of the black hole is sufficient to overcome the star's self gravity and tear it apart \citep{hills75}. A significant fraction of the resulting stellar debris will be accreted onto the SMBH producing a luminous flare. This can be utilised to probe the properties of SMBHs in both active and quiescent galaxies, particularly at modest BH masses, and could be crucial in our understanding of accretion physics, allowing us to observe the formation of accretion discs, the broad line region (BLR) and jets.

High cadence optical surveys have revealed a plethora of nuclear candidates, some with X-ray counterparts and some without \citep{Auchettl2017}. The resulting well-sampled lightcurves and spectroscopic follow-up, while sharing some similarities, have shown a wide range of luminosities, lightcurve shapes and spectral features. The peak luminosities of these transients fall in the range of $L_{\rm bol} \sim 10^{41-44}$ erg s$^{-1}$ and are consistent with blackbody temperatures of $T \sim 20\,000-50\,000$\,K. Their emission fades away over periods of months to years. It is hard to prove that the observed outbursts result from a star being disrupted rather than due to an unusual kind of supernova or an instability in a quiescent AGN disc, but the circumstantial evidence seems strong. Determining the true nature of these flares requires combining detailed observations with theory and simulations.

The lightcurves of TDEs were originally predicted to follow a $t^{-5/3}$ power-law, matching the fall-back rate of material onto the black hole \citep{rees88, phin}. While early discoveries of TDE candidates in X-rays agreed well with this prediction \citep{komassa17}, more recent optical discoveries have revealed a great diversity in lightcurve shapes. This suggests that the mechanism behind the flare is not fully understood. It is widely accepted that one of two scenarios dominate the UV+optical emission. One is an accretion disc which forms from the bound debris. In this scenario, a disc forms as material circularises and a hot blue continuum is emitted peaking in the UV/X-ray. However, observed temperatures are too low to originate from a compact accretion disc, a problem also seen in AGN \citep{andy12}. To explain this and the appearance of broad emission lines, some of the emission would have to be reprocessed. \citet{strub09} suggested unbound material could be responsible for the reprocessing. However, \citet{guill14} showed that self-gravity would make the stream of unbound debris thin and thus not likely to reprocess much emission, though material that was initially bound but later unbound due to winds might still be responsible for some or all of the reprocessing. Another source of reprocessing could be a photosphere surrounding the disc (see also \citet{roth}). In addition to the temperature problem, time integrated energies are more than an order of magnitude lower than expected \citep{piran15}, which could be explained by a significant amount of emission at unobservable EUV wavelengths, by severe mass loss during circularisation \citep{metstone16} or by a long tail in the lightcurve \citep{vv19}. Another issue with the accretion disc scenario is that, with some notable exceptions (e.g. \citealt{15oi}), observations of TDEs have shown the temperature to remain roughly constant as the luminosity declines, when one would expect the temperature to evolve as the disc size or accretion rate changes. 

An alternative process is emission produced by shocks in collisions between stellar debris streams \citep{piran15,shiokawa15,stream,bonnerot17}, though this struggles to produce both X-ray and optical emission simultaneously and also suffers from the missing energy problem. The latter is usually explained by a low radiative efficiency, which \citet{svirski17} explain could be the result of an elliptical accretion disc. Light curve plateaus and late-time X-ray emission have been observed in some TDEs (e.g., \citealt{fyk, gezari17}). These may be due to delayed accretion disc formation resulting from the stream-stream collisions. If so, one might expect a scenario in which shocks in stream-stream collisions dominate the emission initially, then at later times reprocessed emission from the accretion disc takes over. \citet{dai} suggest a model analogous to the unified model of AGN. Here the successful detection of X-ray emission depends on viewing angle as emission in the accretion disc plane gets obscured. The viewing angle may also influence the features seen in the optical spectra \citep{eqx}.

The emission lines featuring in optical TDE spectra show broad ($1-2\times10^4$ km s$^{-1}$; \citealt{arcavi14}), complex, often asymmetric profiles. The spectra of the first optical/UV TDE featured only He~II $\lambda4686$ \citep{gez12} but the ever-increasing sample of events have also shown H~I Balmer and He~I lines, revealing a continuum of H-rich to He-rich TDEs \citep{arcavi14}. In some cases emission lines have been observed to appear and disappear as the TDE evolves \citep{ps18kh2, eqx}. In addition, metal lines such as Fe~II, O~III and N~III have been detected in a number of TDEs. Fe~II can be associated with the formation of a compact accretion disc \citep{fyk}, while O~III and N~III are attributed to Bowen fluorescence, a mechanism whereby an EUV photon is emitted by recombining He~II which resonates with an O~III transition, which in turn resonates with a transition in N~III \citep{blag19, lelo19, onori19}. The detection of Bowen lines in X-ray quiet TDEs supports the existence of an obscured EUV/X-ray emission source. \citet{vv20} find that TDEs showing Bowen lines have smaller radii and a longer rise time in their lightcurves, putting them in a separate class to other TDEs.

\begin{figure}
\includegraphics[width=\columnwidth]{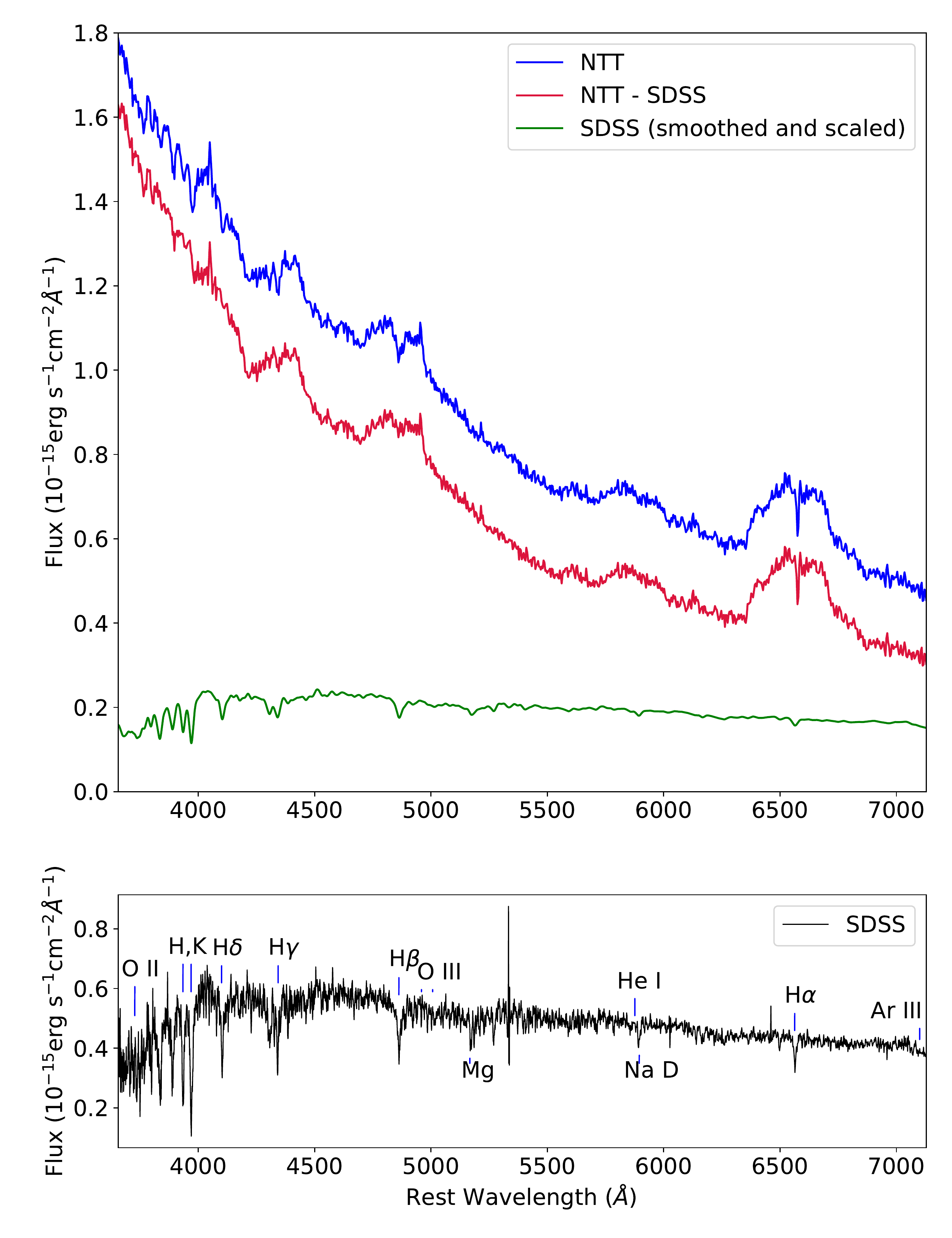}
\caption{Lower: The original SDSS spectrum. Upper: (green) the SDSS spectrum degraded to match the resolution and scaled to the slit width of the NTT spectra, (blue) an NTT spectrum from 02/12/2018 and (red) the same NTT spectrum minus the degraded and scaled SDSS spectrum.
}
\label{fig:hostsub}
\end{figure}

\begin{figure*}
\centering
\includegraphics[width=0.9\linewidth,trim={0 2cm 0 3cm},clip]{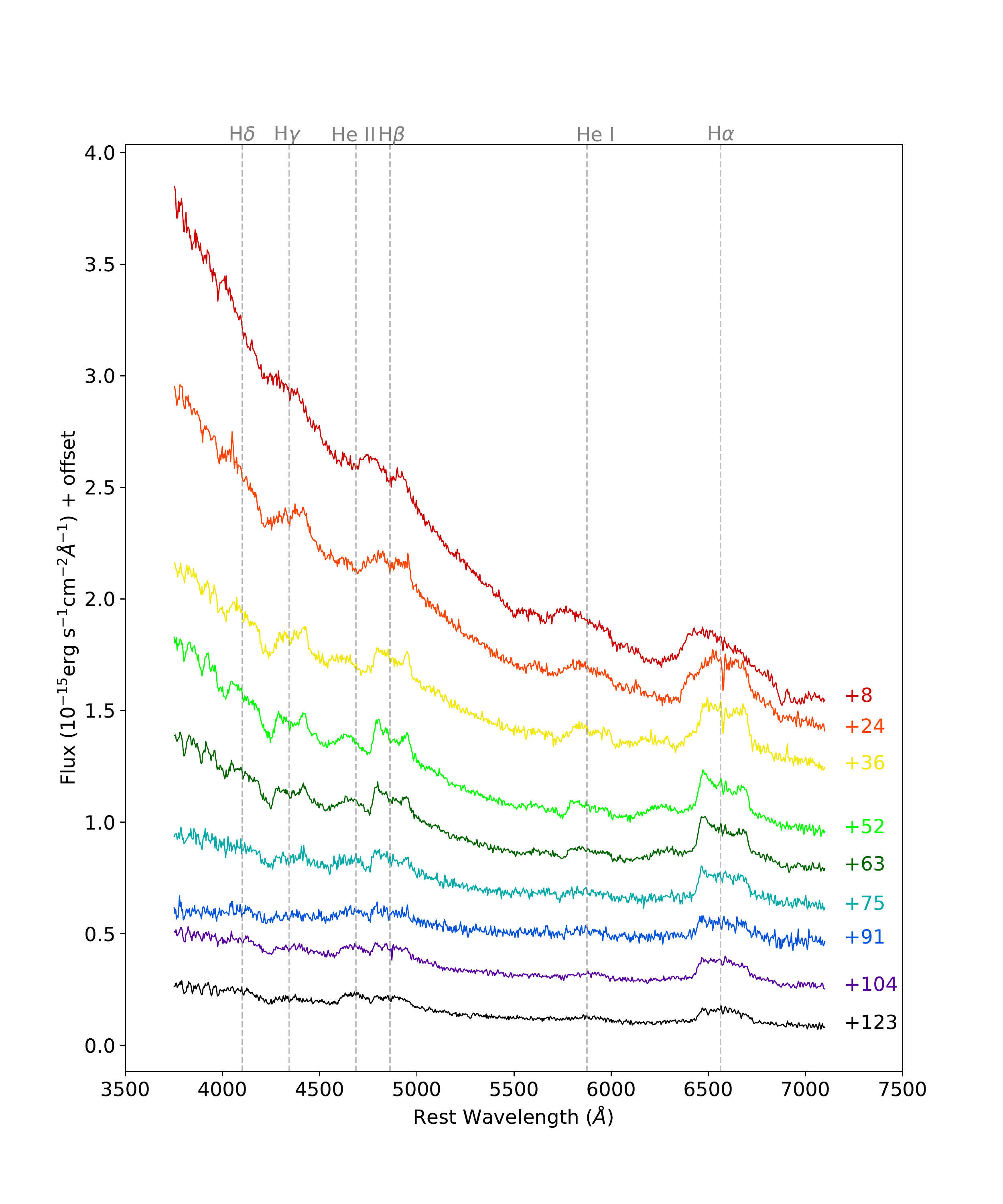}
\caption{Host-subtracted NTT spectra up to 123 rest-frame days from peak. Features in later spectra are too faint to be visible on this scale. The spectra have been offset for clarity. The number on the right marks how many rest-frame days after peak the spectra were taken.
}
\label{fig:host_subbed}
\end{figure*}

\begin{figure*}
\centering
\includegraphics[width=\linewidth,trim={2cm 0cm 2cm 1cm},clip]{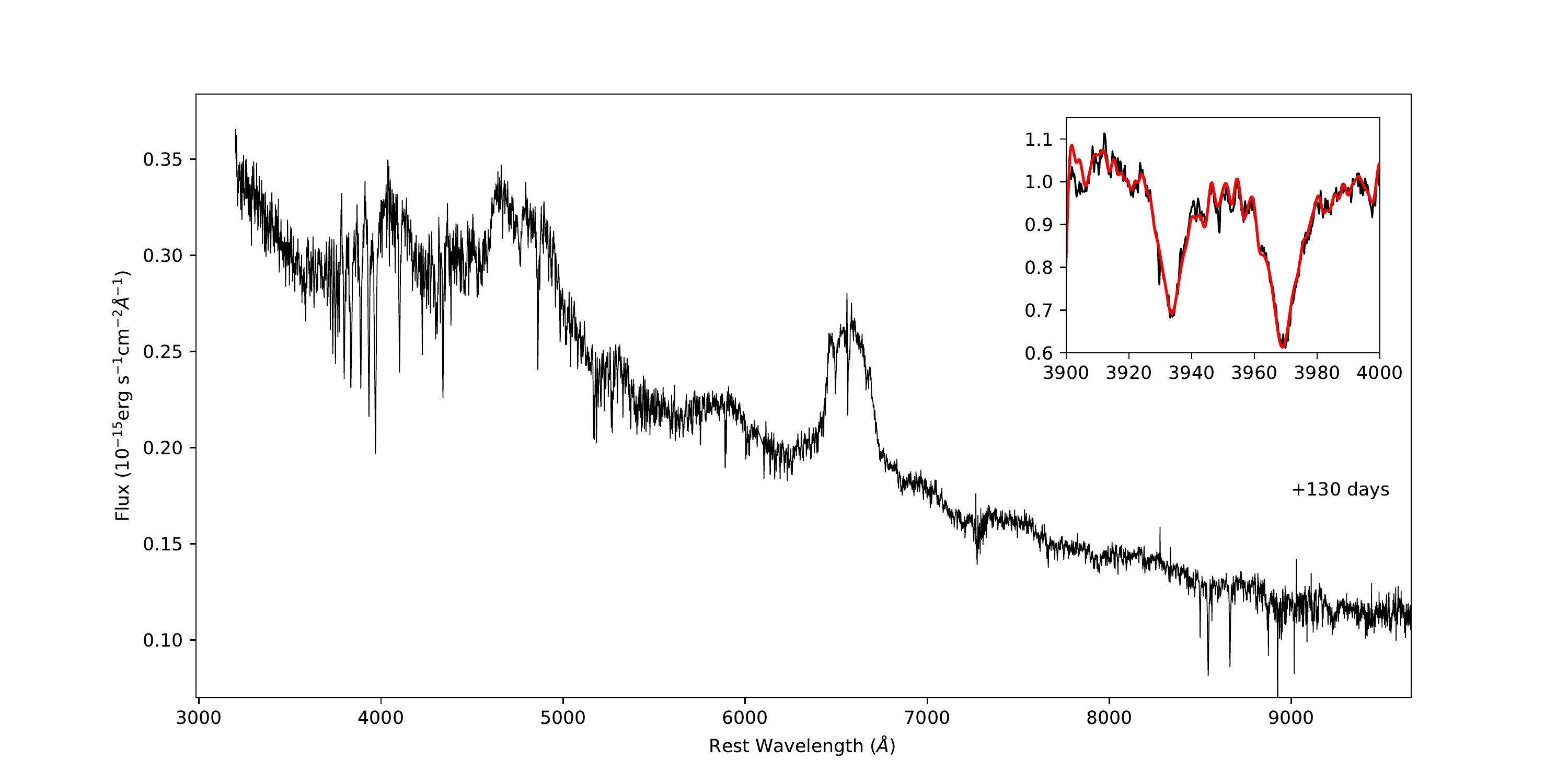}
\caption{The X-Shooter spectrum of AT 2018hyz. Insert: The normalised X-Shooter spectrum (black) zoomed in on Calcium H and K absorption lines. The best fitting model is overlaid in red.}
\label{fig:xshoot}
\end{figure*}

The rich phenomenology of objects classified as TDEs is both fascinating and confusing, and the problems noted -- cool temperature and low total energy -- are a challenge to theoretical understanding. In this paper we present densely sampled spectroscopic observations of the TDE candidate AT~2018hyz (ASASSN-18zj), discovered by the All-Sky Automated Survey for Supernova (ASAS-SN; \citealt{ass1, ass2}) and appearing also in the sample of TDEs detected by the Zwicky Transient Facility (ZTF; \citealt{vv20}). In \S\ref{observations} we present the host galaxy spectrum and follow-up observations, in \S\ref{analysis} we determine the host black hole mass and analyse emission lines from the transient, in \S\ref{discussion} we discuss the results and offer our interpretation and in \S\ref{conclusion} we summarise our findings and come to a conclusion. A companion paper, \citet{gomez}, discusses photometry, key results from which we use in this paper\ref{hung}. \citet{gomez} find the peak of the bolometric lightcurve to be at ${\rm MJD} = 58429$ so for consistency we define the epoch of each spectrum relative to this. \footnote{\label{hung} While this paper was under review, another paper on the same event appeared on arXiv by \citet{hung20} which we include in our discussions.}

\begin{figure*}
\centering
\includegraphics[width=1\linewidth,trim={2cm 3.5cm 2cm 5cm},clip]{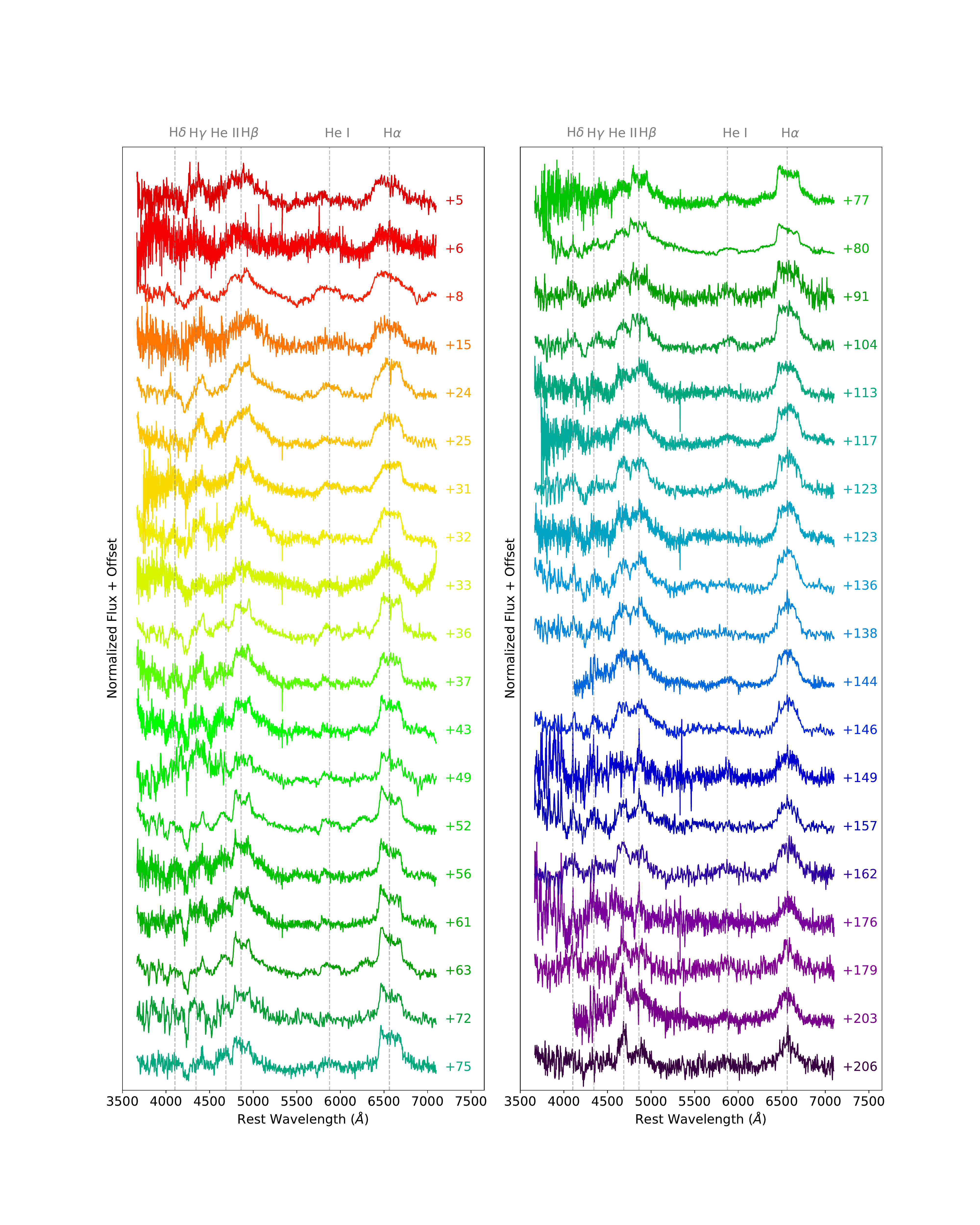}
\caption{All host and continuum subtracted spectra. The days after peak in the TDE rest frame are marked on the right.}
\label{fig:cont_subbed}
\end{figure*}

\begin{figure*}
\includegraphics[width=0.9\linewidth,trim={1cm 0cm 1cm 0cm},clip]{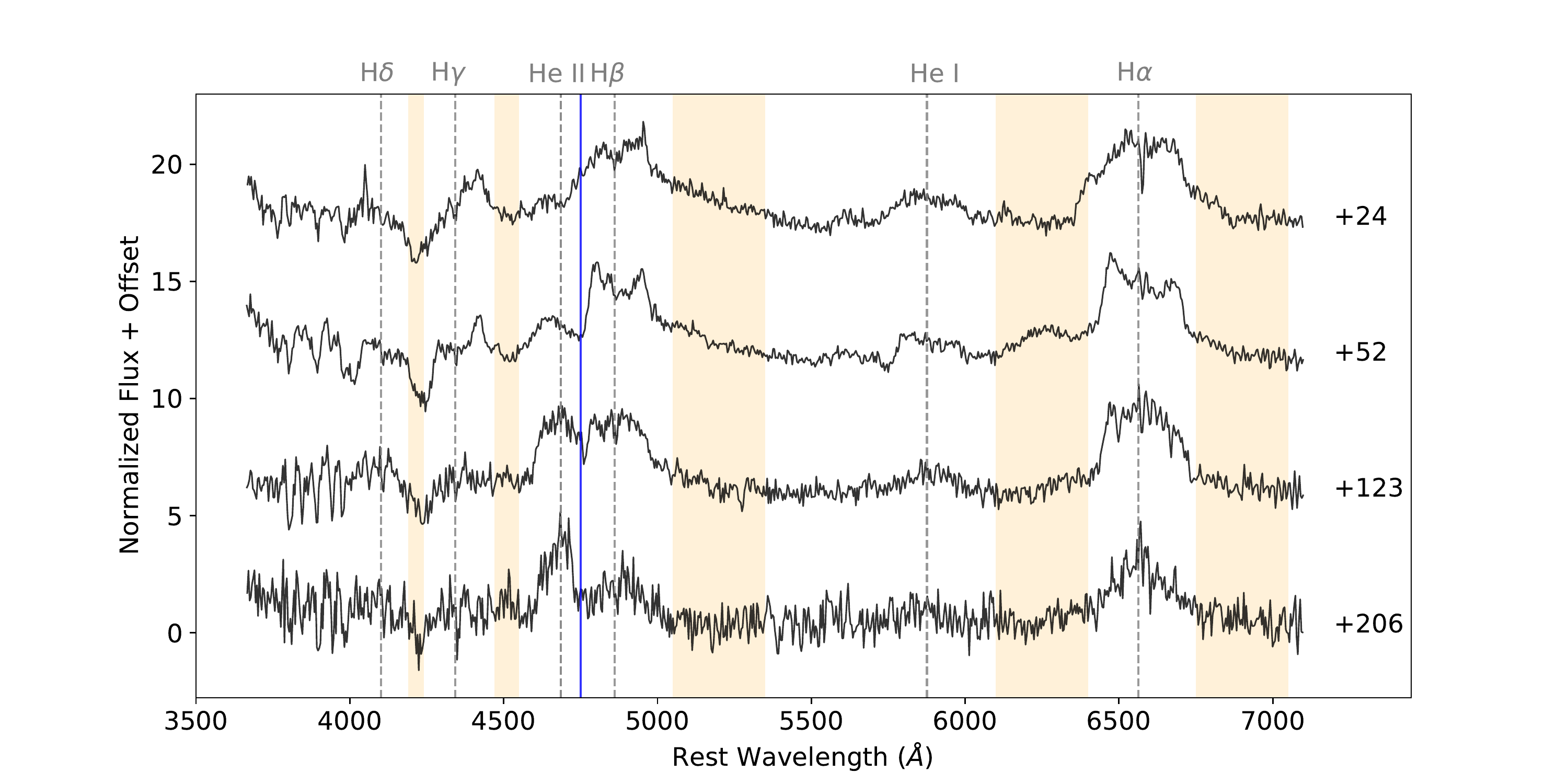}
\caption{Continuum subtracted spectra from different epochs with emission lines labelled. The time after peak (in rest-frame days) when each spectrum was taken is marked on the right. The regions used in local continuum subtractions are highlighted in orange. After continuum subtraction the area between these regions was integrated to find the line flux, with the exception of H$\beta$, where the lower limit is the solid blue line to avoid He II.
}
\label{fig:lab}
\end{figure*}

\begin{figure}
\includegraphics[width=\columnwidth]{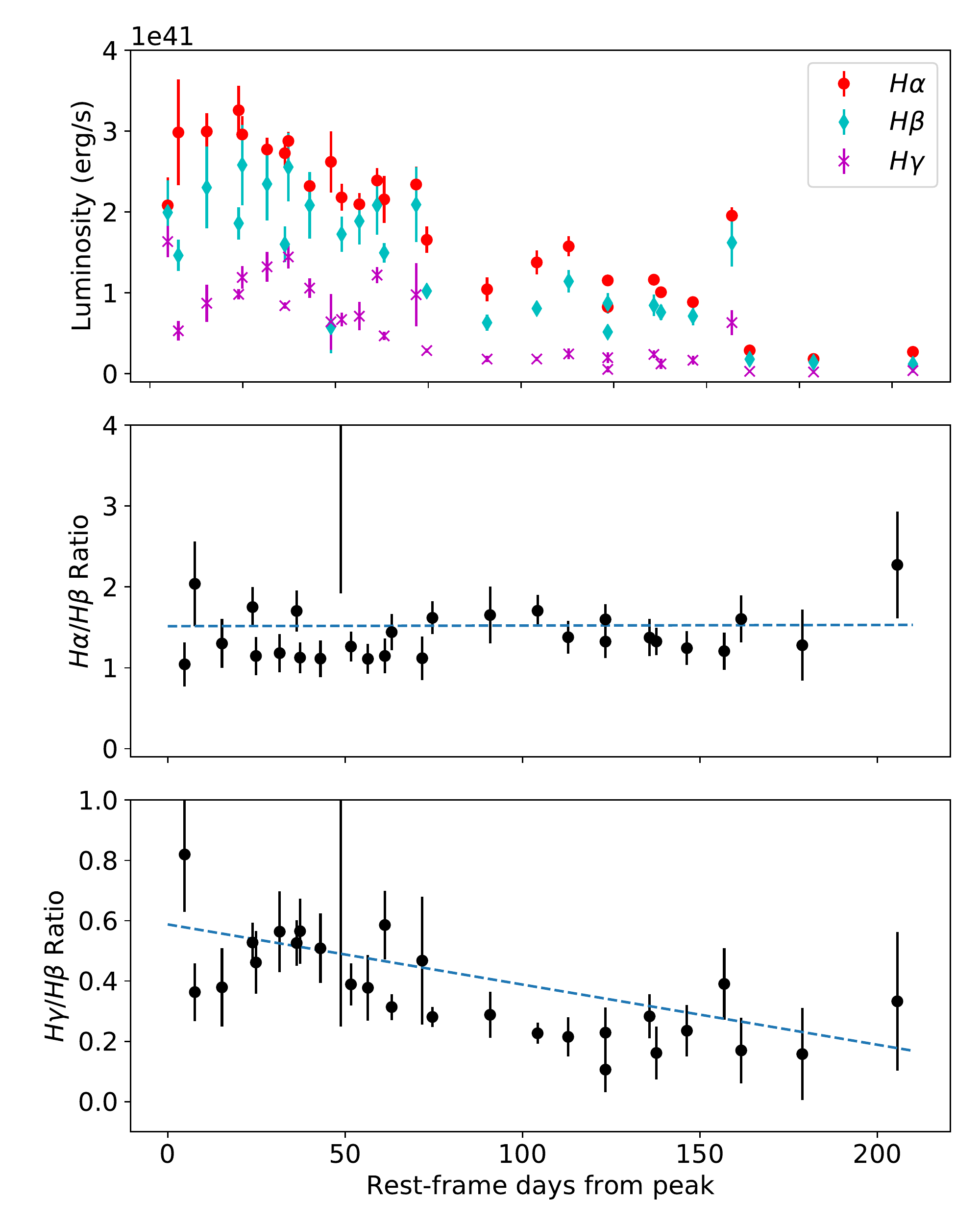}
\caption{The H$\alpha$, H$\beta$ and H$\gamma$ luminosities fade with time (top), H$\alpha$/H$\beta$ stays roughly constant (middle), while H$\gamma$/H$\beta$ appears to decline. The dashed lines shows the line of best fit. The error bars are underestimated as they do not include systematics.
}
\label{fig:ratios}
\end{figure}

\begin{figure*}
\includegraphics[width=\linewidth,trim={2cm 1cm 3cm 2cm},clip]{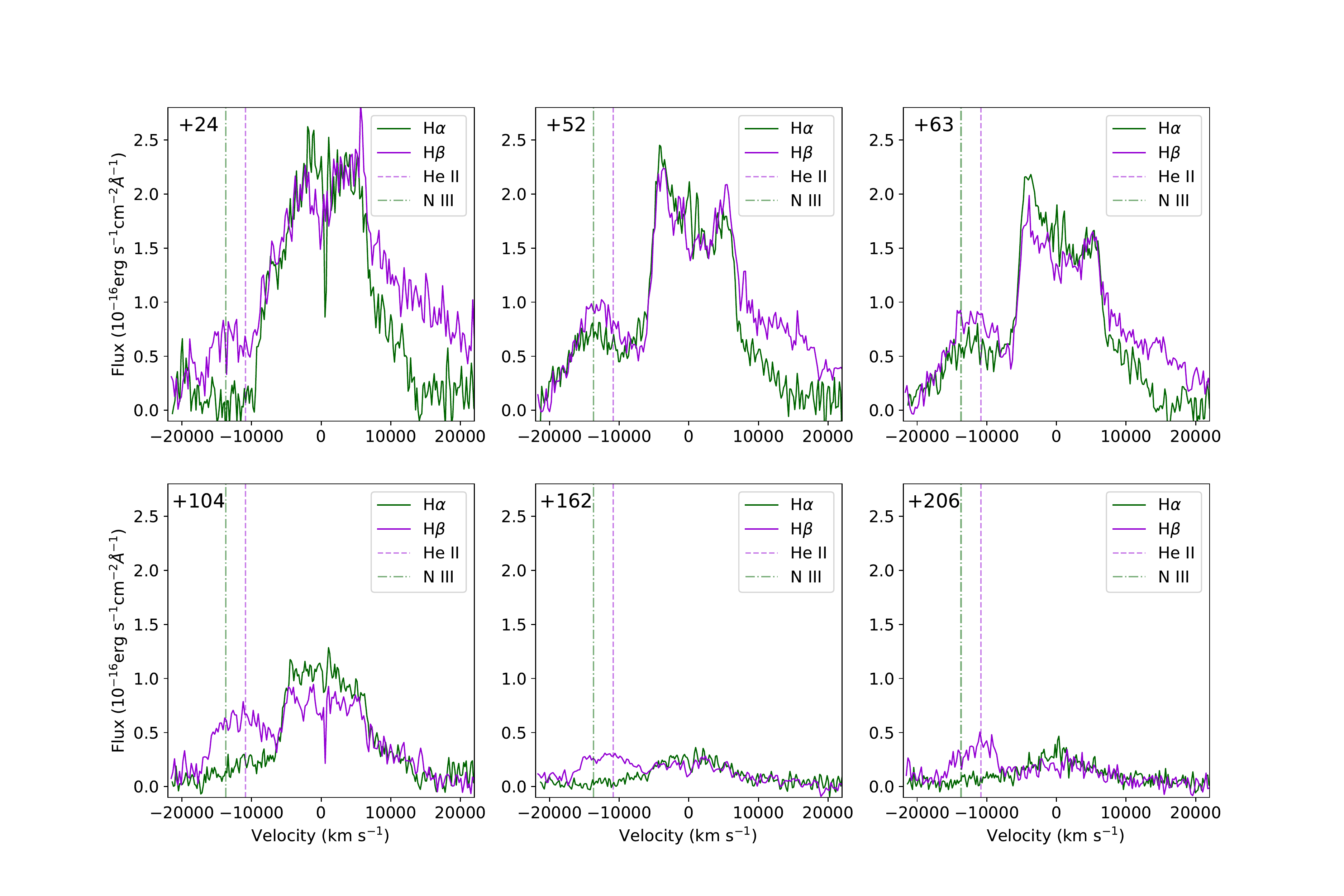}
\caption{Velocity profiles of H$\alpha$ and H$\beta$ from six epochs (marked in the top left in rest-frame days from peak). At +52 days, the blue bump in Ha is at a similar velocity to what we had assumed to be He II, possibly suggesting a blue-shifted outflow ($\sim$ 12000~km~s$^{-1}$). At +162 days a second peak appears on He II that matches the position of the N III line produced by Bowen fluorescence.}
\label{fig:vel}
\end{figure*}

\section{Observations and Data Reduction} \label{observations}

AT~2018hyz was discovered by ASAS-SN on 2018-11-06.65 (all times in UT) with an apparent $g$-band magnitude of 16.4 mag \citep{atel1}, though the earliest ASAS-SN detection of this event was on 2018-10-14.48. The transient was detected at $\alpha$ = 10:06:50.872, $\delta$ = +01:41:34.10, placing it $0.2\pm0.8$" \citep{gomez} from the centre of \textit{SDSS J100650.86+014134.0}, an apparently quiescent or post-starburst galaxy at a redshift of $z = 0.04573$. Assuming a flat $\Lambda$CDM cosmology with \mbox{$H_{0} = 69.3$ km s$^{-1}$ Mpc$^{-1}$}, $\Omega_{m} = 0.286$, and $\Omega_{\Lambda} = 0.712$ \citep{Hinshaw13}, the redshift of the host galaxy suggests an absolute $g$-band magnitude of $-20.2$ mag. AT\,2018hyz was classified as a TDE candidate following spectroscopic observations \citep{atel2} which show a strong blue continuum with broad Balmer lines and He I, similar to other TDE candidates (e.g. \citealt{arcavi14, 14ae, ps18kh}). The combination of these spectral features, high luminosity and proximity to the nucleus of the host make AT\,2018hyz a strong TDE candidate.

\subsection{Archival host galaxy observations}
An archival spectrum of the host galaxy was available from the Sloan Digital Sky Survey (SDSS; \citealt{sdss}). The spectrum is displayed in Figure \ref{fig:hostsub}. The lack of common emission lines such as H$\alpha$ and [O II], combined with strong Balmer absorptions, indicate that this is a quiescent, Balmer-strong galaxy, otherwise known as a post-starburst or E+A galaxy (i.e. an elliptical galaxy showing absorption lines from A-type star atmospheres). TDEs are over-represented in such galaxies by a factor of $30-35$ \citep{arcavi14,french16,graur18}, possibly because these galaxies have higher central stellar densities than the background population \citep{graur18,lawsmith17}. A stellar population synthesis model is fit to archival host photometry in \citet{gomez}.

\subsection{Spectroscopy}
Follow-up spectra were obtained via the extended Public ESO Spectroscopic Survey of Transient Objects (ePESSTO/ePESSTO+; \citealt{pessto}) with EFOSC2 on the 3.6m New Technology Telescope (NTT) at the La Silla Observatory, Chile. A slit width of 1.0" or 1.5", depending on seeing, was placed on the target at the parallactic angle. Spectra were reduced and calibrated with standard \textsc{iraf} routines via the PESSTO pipeline \citep{pessto}. Flux calibration was achieved using standard star observations obtained on the same nights. 

Las Cumbres spectra were taken with the FLOYDS spectrographs mounted on the 2m Faulkes Telescope North and South at Haleakala (USA) and Siding Spring (Australia), respectively. A 2.0" wide slit was placed on the target at the parallactic angle. One-dimensional spectra were extracted, reduced, and calibrated following standard procedures using the FLOYDS pipeline\footnote{\href{https://github.com/svalenti/FLOYDS_pipeline}{https://github.com/svalenti/FLOYDS\_pipeline}} \citep{val2}. 

Further spectra were taken with the IMACS \citep{dressler11} and LDSS3c spectrographs on the 6.5m Magellan telescopes, OSMOS on the 2.4m Hiltner Telescope at MDM observatory \citep{martini11}, the Boller and Chivens spectrograph on the Ir\'en\'ee du Pont Telescope and the FAST spectrograph on the 1.5m Tillinghast Telescope \citep{fast}. The spectra were processed using standard \textsc{iraf} routines with the \textsc{twodspec} package. Bias and flat field corrections were applied, and the sky background was modeled and subtracted from each image. The one-dimensional spectra were optimally extracted, weighing by the inverse variance of the data. A wavelength calibration was applied using a HeNeAr lamp spectrum taken near the time of each science image. Relative flux calibration was applied to each spectrum using a standard star taken on the same night.

One spectrum was observed with the Intermediate-dispersion Spectrograph and Imaging System (ISIS; \citealt{Jorden1990}) spectrograph at the William Herschel Telescope (WHT) on 6 January 2020. We used a slit width of 1.0" in combination with the R600 gratings. The use of a dichroic results in blue and red spectra, covering $\sim$1400\AA{} around central wavelengths of 4500 and 7000 \AA{} respectively. After performing standard data reduction tasks in \textsc{iraf}, including a bias subtraction and flatfield normalisation, cosmic rays are removed using the \textit{lacos} package \citep{vandokkum2012}. A wavelength calibration is applied using CuAr+CuNe arc lamp frames; the dispersion in the wavelength solution is $<$0.1 \AA. A summary of all observations is shown in Table \ref{tab:obs} in Appendix A.

All spectra up to +206 days after peak were scaled to Las Cumbres photometry, then redshift and extinction corrected using a redshift of $z=0.04573$ and an extinction value of $E(B-V)=0.0288$ mag \citep{ebv} in \textsc{iraf}. We removed host galaxy light by subtracting the archival SDSS spectrum. SDSS spectra have higher resolution (3\AA/pix) than a lot of our data (see Table \ref{tab:obs}). In addition, the host spectrum has a slit width of 3.0" so had to be scaled to match the slit widths of the other spectra. This was performed by multiplying by the fraction of the host galaxy that would appear in each slit using the galaxy half-light radius of 1" obtained from SDSS. After correcting for the different slit widths and scaling to SDSS photometry, we subtracted the host galaxy spectrum from our spectra using the \textsc{sarith} task within \textsc{iraf}. The host-subtracted spectra were then scaled to host-subtracted photometry from \citet{gomez}. This method is illustrated in Figure \ref{fig:hostsub}, while selected host-subtracted NTT spectra are displayed in Figure \ref{fig:host_subbed}. We assumed no host extinction in these subtractions as we see no Sodium absorption. If host extinction is present, it has the effect of dimming AT\,2018hyz and making it appear cooler.

We also obtained spectra with the X-shooter spectrograph on the Very Large Telescope (VLT) in slit-nodding mode . X-shooter is an echelle spectrograph with three arms (UVB, VIS, NIR), with a combined wavelength range from 3200 - 24\,700\,\AA{} \citep{xshoot}. We reduced the spectra with the dedicated \textsc{esoreflex} pipeline (v. 3.3.4). A telluric standard was obtained at a similar airmass to the target, and a telluric spectrum was then generated by dividing the reduced standard spectrum by an interpolated spectrum using the telluric-free regions. The science spectrum was then divided by the residual telluric spectrum to remove these features from the data.

\section{Spectroscopic Analysis} \label{analysis}

\subsection{Host Black Hole Mass}

With the high resolution spectrum from X-shooter displayed in Figure \ref{fig:xshoot}, we were able to accurately measure the host galaxy absorption lines to estimate the central black hole mass via the $M_{\rm BH}-\sigma_{\star}$ relation. We first resampled the spectrum onto a linear dispersion in the logarithm of wavelength, before normalising it to the continuum using 3$^{rd}$ order cubic splines. We masked prominent absorption features, while including TDE emission features to help with comparison to stellar templates.

We then use \textsc{ppxf} \citep{cappellari2017} in combination with the Elodie stellar template library \citep{prugniel2001, prugniel2007} to measure the velocity dispersion using the myriad of stellar absorption features present in the spectrum (see \citealt{wevers2017} for more details). We measure a stellar velocity dispersion of $\sigma_{\star}=57 \pm 1$~km~s$^{-1}$; the best fit template overlaid on the X-shooter spectrum is inset in Figure \ref{fig:xshoot}. Similar measurements were made using the host SDSS spectrum and our WHT spectrum, giving $\sigma_{\star}=59 \pm 3$~km~s$^{-1}$ and $58\pm 5$~km~s$^{-1}$ respectively, consistent with the X-shooter result.

Using the $M_{\rm BH}-\sigma_{\star}$ relation of \citet{mcma}, the measured velocity dispersion corresponds to a black hole mass of log($M_{\rm BH}/M_{\odot}) = 5.25 \pm 0.39$ where the quoted uncertainty includes both the systematic and statistical error. Such a low black hole mass is consistent with the lowest mass black holes in the sample of known TDEs in \citet{wevers2019}, though we note that our low measurement of $\sigma$ is significantly out of the range covered by the data used to derive the McConnell \& Ma relation. In addition, TDEs typically have an Eddington ratio of 0.1-1 at peak \citep{wevers2017,Mockler2019}, whereas the luminosity of AT\,2018hyz at peak, $1.9\times10^{44}$~erg~s$^{-1}$ \citep{gomez}, would correspond to $\sim10 L_{\rm Edd}$ for this black hole mass. 

If we instead use the $M_{\rm BH}-\sigma_{\star}$ relation of \citet{koho}, we obtain log($M_{\rm BH}/M_{\odot}) = 6.09 \pm 0.30$. This is more consistent with the peak luminosity being at $L_{\rm Edd}$. Although this value is closer to what we expect, this relation is derived exclusively for classical elliptical bulges which may or may not apply in this case. \citet{gomez} find log($M_{\rm BH}/M_{\odot}) = 6.72\pm0.20$ from lightcurve modelling, which is higher than we measure from the spectra. Similarly, \citet{hung20} measure log($M_{\rm BH}/M_{\odot}) = 6.55^{+0.17}_{-0.13}$ from lightcurve modelling and log($M_{\rm BH}/M_{\odot}) = 6.2$ from a different $M_{\rm BH}-\sigma_{\star}$ relation. They are in agreement that the \citet{mcma} relation gives an improbably low mass. We add a further mass estimate using the $M_{\rm BH}-M_{\rm bulge}$ relation from \citet{mcma}. \citet{Mendel14} find a bulge mass of $\log{(M_{\rm bulge} / M_\odot)} \sim 8.8$, which gives a black hole mass of log($M_{\rm BH}/M_{\odot}) \sim 6.2$ consistent with our other estimates. As it is not clear which measurement should be favoured, it seems reasonable to assume a black hole mass of $10^6$M$_{\odot}$ for the rest of this analysis, with a large uncertainty of a factor $\gtrsim 2$, which gives a peak Eddington ratio of $\sim1.5$. 

\subsection{Transient Line Evolution}

The spectra of AT 2018hyz are dominated by a blue continuum. \citet{gomez} measure the black-body continuum temperature from photometry, which drops from an initial temperature of $\sim$22\,000 to 16\,000\,K after 50 days, before rising back up to 21\,000\,K. To isolate and study the spectral line evolution, we first removed this continuum. To produce the spectra shown in Figure \ref{fig:cont_subbed}, the continua were fit using 4th-order polynomials with the \textsc{IRAF} continuum package, before being subtracted from the spectra. The subtractions are satisfactory for line visualization, but over-subtract at shorter wavelengths especially at earlier times.

The spectra initially show broad Balmer lines and He~I~$\lambda5876$. At around 70-100 days after peak, He~II~$\lambda4686$ appears and increases in strength. The Balmer line profiles are complex and evolve dramatically. The profile is initially roughly Gaussian but slightly boxy and asymmetric. It becomes boxier with time and develops a double peaked shape $\sim40$ days after discovery with a stronger blue peak than red, maintaining the asymmetry. The profile loses the double-horned shape around 90 days after discovery, becoming flat topped before rounding off, losing the asymmetry and developing a Gaussian or maybe even a triangular profile. He I is weaker but appears to follow a similar evolution, suggesting it arises from a similar physical region. In some spectra (e.g. phase $24-75$), H$\gamma$ appears redshifted. This could be the red peak of a double-peaked line, but we would also expect to see the blue peak. It is not clear what the cause of this feature is. We measured the FWHM of H$\alpha$ in each spectrum empirically, as a Gaussian does not consistently provide a good fit. The profile narrows over the period of observations from $\sim$17\,000~km~s$^{-1}$ to $\sim$10\,000~km~s$^{-1}$. If the observed velocity width corresponds to Keplerian rotation of a disc, an FWHM of 17\,000~kms~$^{-1}$ implies the orbital radius of the emitting material is $R\sim600\,R_{\rm s}$, whereas 10\,000~km~s$^{-1}$ corresponds to a radius of $R\sim1800\,R_{\rm s}$ (where $R_{\rm s}$ is the Schwarzschild radius, which for a $10^6$~M$_{\odot}$ black hole is $\sim0.02$~AU), though these values would be lower for an inclined disc. The initial high velocity is measured before the profile looks disc-like, hence the velocity width may not correspond to a meaningful rotational velocity.

\begin{figure*}
\includegraphics[width=\linewidth,trim={1.5cm 0.5cm 1.5cm 1cm},clip]{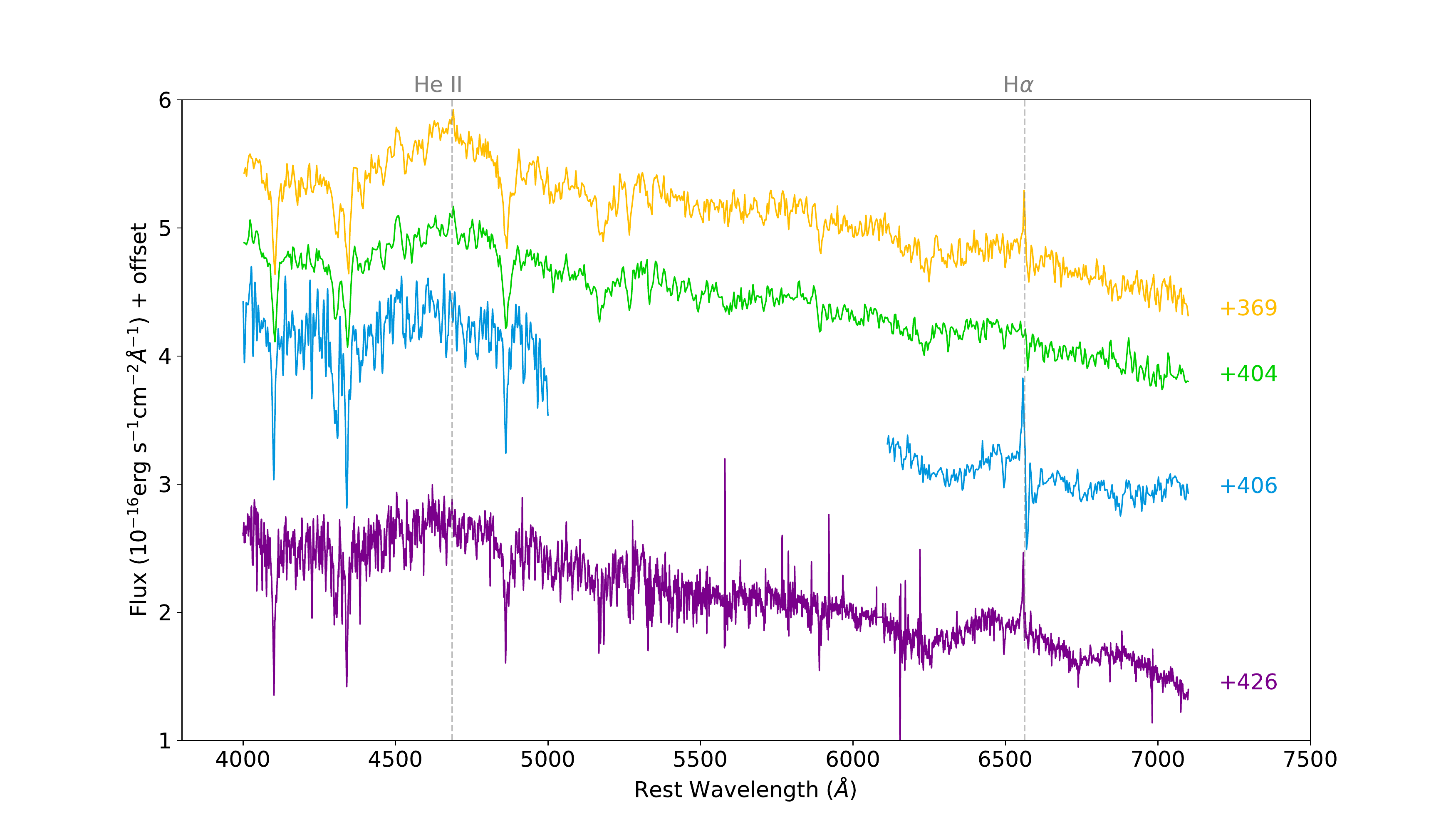}
\caption{Late-time spectra taken once AT\,2018hyz was no longer obscured by the Sun. The first two are from the NTT, the third from the WHT and the most recent is from X-Shooter. The host has not been subtracted from these spectra. A narrow H$\alpha$ line has appeared and He II is just visible in the earliest of these spectra
}
\label{fig:newspecs}
\end{figure*}

\begin{figure}
\includegraphics[width=\columnwidth,trim={0.1cm 0.3cm 0.2cm 0.5cm},clip]{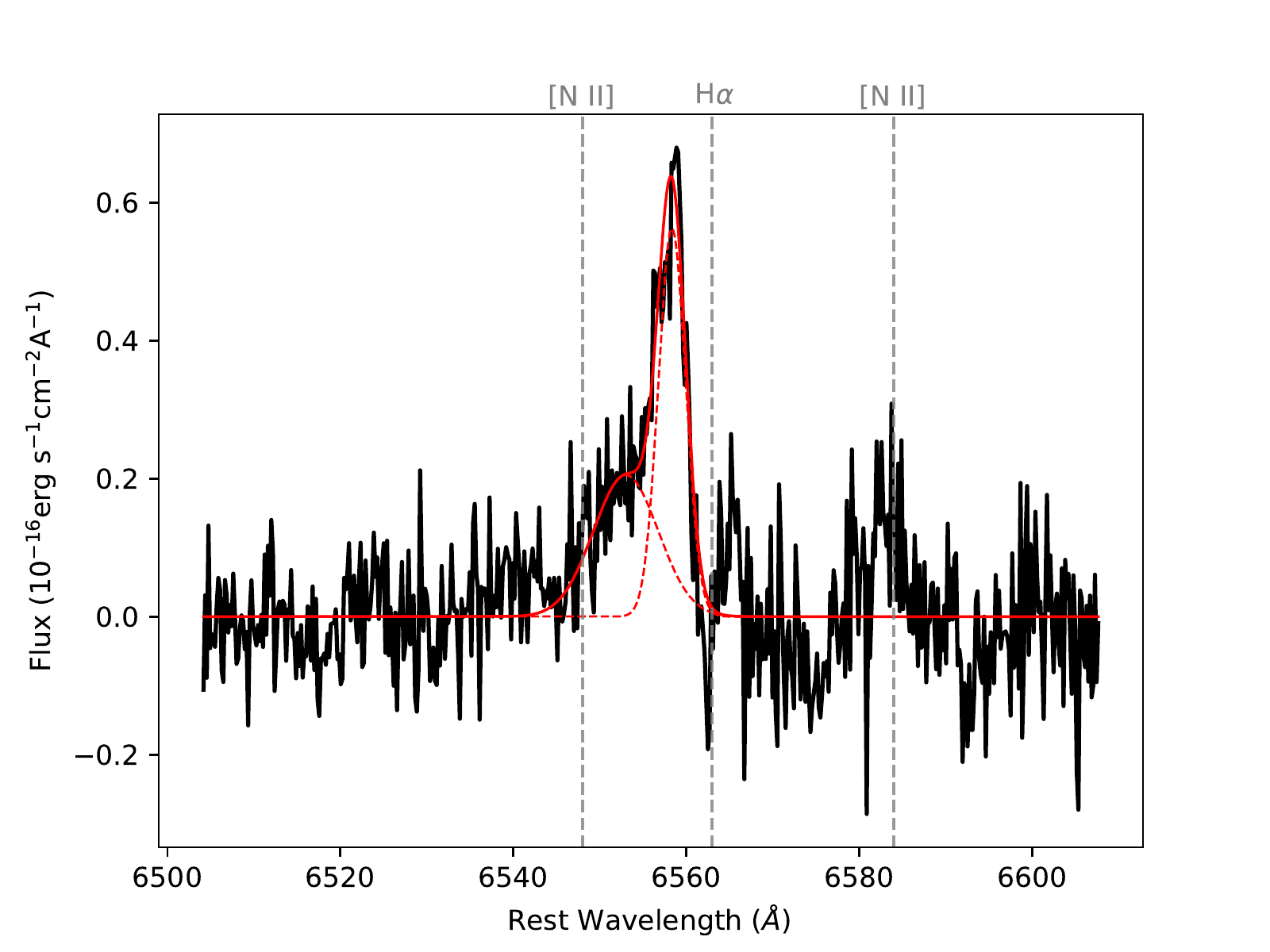}
\caption{A two Gausian component fit to the narrow H$\alpha$ line in the X-Shooter spectrum taken +426 days from peak. The line appears blueshifted, with the narrow component centered on 6558\AA{}.
The broad component might be [N~II]~$\lambda$6548, with the corresponding [N~II]~$\lambda$6584 to the right.
}
\label{fig:hafit}
\end{figure}

\begin{figure*}
\centering
\includegraphics[width=0.9\linewidth,trim={0cm 0cm 0cm 0cm},clip]{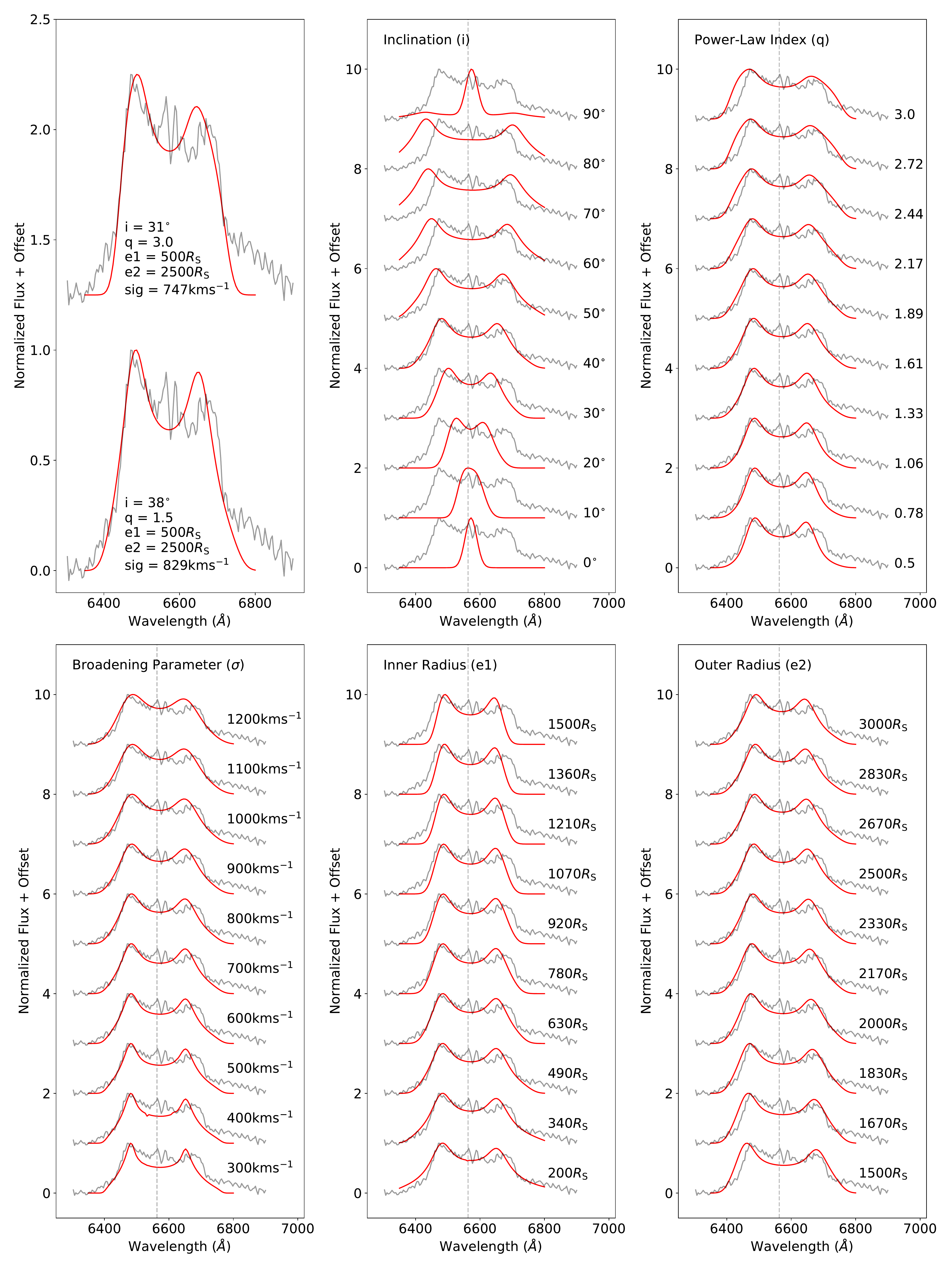}
\caption{The top left panel shows examples of best fitting models to our spectrum. The remaining panels explore parameter space of the circular disc model. In each of these plots only one parameter is varied, while the others are fixed at $i=38^{\circ}$, $q=1.5$, $\sigma = 500$~km~s$^{-1}$, $e_1=500R_{\rm S}$ and $e_2=2500R_{\rm S}$.}
\label{fig:discmod}
\end{figure*}

Once the He~II line becomes clearly visible, it is well fit by a Gaussian with a FWHM of 7700~km~s$^{-1}$, corresponding to Keplerian rotation at $\sim3000\,R_{\rm s}$. However, at +162 days we observe a secondary peak just blueward of He II which could be due to the N~III~$\lambda4640$ line, suggesting the presence of Bowen fluorescence. We fit a double Gaussian, one at 4640\,\AA{} and one at 4686\,\AA{}, allowing the widths to vary independently. The N~III and He~II lines are fit by Gaussians with FWHMs of 3800~km~s$^{-1}$ and 4200~km~s$^{-1}$ respectively. This suggests that, if it is contaminated by Bowen fluorescence, He II originates further out at $\sim10\,000\,R_{\rm s}$, assuming the gas is virial. \citet{vv20} show that TDEs with Bowen lines tend to have black-body radii smaller than $6\times10^{14}$cm; the size of the black-body radius inferred for AT2018hyz by \citet{gomez} is initially $1.3\times10^{15}$cm suggesting we should not expect Bowen fluorescence in AT\,2018hyz. However, the radius shrinks over time and reaches $6\times10^{14}$cm after $\sim100$ days, so it may be possible for these lines to form then. The appearance of He~II and weakening of H$\alpha$ is somewhat similar to AT~2017eqx \citep{eqx}, though in the latter case He~II is always strong and H$\alpha$ completely disappears. The change in line strengths and their appearance/disappearance suggests that it may be possible for events to transition between the TDE classes (H, He and Bowen) defined by \citet{vv20}.

Measuring the line flux was not trivial due to blending between neighbouring broad features. To obtain more reliable measurements we used local continuum subtractions. For each of H$\alpha$, H$\beta$ and H$\gamma$, we defined relatively line-free wavelength ranges on each side of the emission profiles. The lower and upper ranges, respectively, were $6100-6400$\,\AA{} and $6750-7050$\,\AA{} for H$\alpha$, $4470-4550$\,\AA{} and $5050-5350$\,\AA{} for H$\beta$ and $4190-4240$\,AA{} and $4470-4550$\,\AA{} for H$\gamma$. Within these ranges, we randomly selected regions of the spectra and fit a linear continuum, which was subtracted before integrating the line flux. For each line and each spectrum, this process was iterated 1000 times to determine a mean flux and standard deviation, which we take as an estimate of the uncertainty. The regions are highlighted in Figure \ref{fig:lab}, where it is clear how difficult it is to select a region with no contamination. As shown in Figure \ref{fig:ratios}, the H$\alpha$, H$\beta$ and H$\gamma$ emission lines decay over time, H$\alpha$/H$\beta$ stays at $\sim$1.5, while H$\gamma$/H$\beta$ drops from $\sim0.6$ to $\sim0.2$, though our H$\gamma$ measurement is less reliable as it is a weak line in a contaminated region. The line fluxes we measure, in particular H$\beta$, are dependent on where we define our continuum regions. However, after experimenting with different regions we find H$\alpha$/H$\beta$ is always below 2, and any contamination unaccounted for in the H$\beta$ region would mean the true value is lower. The presence of host-extinction would also make the true value of H$\alpha$/H$\beta$ lower.

Figure \ref{fig:vel} shows H$\alpha$ and H$\beta$ in velocity space over six epochs. The two lines show a similar evolution except for the red shoulder on H$\beta$. We observe a feature appearing at the position of He II 4686\AA{} at $\sim36$ days after discovery. A feature also develops blueward of H$\alpha$ at the same time at around 6300\AA{}, suggesting it is related to the appearance of the 4686\,\AA{} feature.

The order of appearance of the Balmer lines and He II makes sense if we consider the dynamical timescale, $t_{\rm dyn}=\sqrt{R^3/GM}$. From the velocity width of the Balmer lines we estimate the emitting region to initially be at a radius of 600$\,R_{\rm s}$. For a $10^6M_{\odot}$ black hole this gives a dynamical timescale of $\sim2$ days. The radius of the emitting region of the He II line estimated from the line width depends on whether or not the Bowen component is present. Assuming pure He II, the emitting region is at $3000\,R_{\rm s}$ and has a dynamical timescale of $t_{\rm dyn}\sim30$ days. With the Bowen component the line comes from $10\,000\,R_{\rm s}$ and has $t_{\rm dyn}\sim160$ days. While the ordering of appearance of the lines is consistent with their dynamical timescales, the appearance of He II after $\sim70-100$ days is consistent only if the line is at the larger radius, requiring contamination from the Bowen N III line.

After the spectrum 206 rest frame days from peak, AT\,2018hyz became obscured by the Sun. Following its reappearance we obtained a further two spectra with the NTT, one with the WHT and one with X-Shooter on the VLT. These are displayed in Figure \ref{fig:newspecs}. He II is just visible in the earliest of these and a narrow line has appeared at the position of H$\alpha$. At the redshift of AT\,2018hyz, H$\alpha$ falls on a sky line at the edge of the Telluric B band, making its veracity uncertain in the NTT spectra. However, the X-Shooter and WHT spectra are of sufficiently high resolution to separate this feature from the telluric band. We subtract the local continuum around the line from the X-Shooter spectrum and fit a two-component Gaussian, displayed in Figure \ref{fig:hafit}. The broad component is blueshifted by $-460$~km~s$^{-1}$ and has a FWHM of $400$~km~s$^{-1}$, while the narrow component is blueshifted by $-210$~km~s$^{-1}$ and has a FWHM of $170$~km~s$^{-1}$. However, the line profile is likely affected by host-absorption which we are unable to correct for. The velocity widths of the broad and narrow components correspond to radii of $1.1\times10^6R_{\rm S}$ and $6.2\times10^6R_{\rm S}$ respectively. The broad component could be due to [N~II]~$\lambda$6548. It appears redshifted though a feature that could be [N~II]~$\lambda$6584 is also present (see Figure \ref{fig:hafit}). The most intuitive interpretation of the narrow line is that it is a light echo, produced when photons from the TDE interact with a slow moving gas cloud at a large radius. One potential effect of this is that our temperature measurements are higher than they should be, as earlier, hotter spectra contribute to later ones (e.g. as in \citealt{1991T}). The temperature increase observed after +50 days could be a result of this, however the narrow line doesn't appear for around another 200 days so we don't believe the two are related. The light travel time from the black hole to the radius of the narrow component is $\sim700$ days. This is consistent with our observations (the line appears some time between our spectra at +206 and +369 days) if the angle between gas cloud and line-of-sight to the TDE, from the position of the black hole, is in the range $45^{\circ}$ to $60^{\circ}$.

\subsection{Disc model fitting}
We fit models to the double-peaked emission lines using prescriptions for a circular accretion disc described by \citet{chen89} and \citet{Strat03}. We acknowledge that in reality the scenario would likely be more complex than a circular disc, but the purpose of the fitting is to show that a disc can produce the observed profiles. Initially we leave 4 parameters free: the inclination of the disc (\textit{i}), the power-law index (\textit{q}) which controls the emission line wing shape where $R^{-q}$ gives the emissivity profile, the inner line emitting radius ($e_1$) and a local broadening parameter (\textit{$\sigma$}). We keep the outer radius fixed at 2500$R_{\rm S}$ as this has little effect on the resulting line profile. We fit the model to the continuum-subtracted H$\alpha$ region in the spectrum at +52 days, chosen as it shows a clear double peak. Before fitting the disc model, the bump bluewards of H$\alpha$ (Section 3.2) was removed by fitting a Gaussian and subtracting. The parameters of the best fitting models, shown in the top left panel of Figure \ref{fig:discmod}, are $i=30.5^{\circ}$, $q=3$, $e_1=500R_{\rm S}$ and $\sigma=747$~km~s$^{-1}$. These best fitting models also provided good fits to the H$\beta$ emission. We then tried fixing $e_1$ and \textit{q} to allow finer grid spacing for \textit{i} and \textit{$\sigma$}. $e_1$ was fixed at 500$R_{\rm S}$ as this value consistently fit the line well, and \textit{q} was set to 1.5 as this value is found observationally in CV discs \citep{horne91}. The fits both look reasonable and support the idea that the double-peaked profiles can be interpreted as evidence of an accretion disc. However, at early and late times (when the Balmer lines do not show the double-peaked profile) the disc model fails to provide a good fit, suggesting that either the emission is being obscured/reprocessed, or the structure of the disc itself is changing. We show how varying different parameters alters the emission line shape in the remaining panels in Figure \ref{fig:discmod}.

\citet{hung20} perform emission line fits to multiple epochs using an elliptical disc model combined with a Gaussian component. They find that while the Gaussian component dominates in later epochs, the double peaked profile is still required for a good fit, suggesting the disc is never completely obscured but that the emission becomes dominated by a radiatively-driven wind. The velocity dispersion fit by \citet{hung20} is lower than the value we obtain and the inclination is higher, though both are within $2\sigma$ of our estimates. This is likely due to the Gaussian component fitting the broader wings thus allowing the disc model to only fit the double peaks. The combination of elliptical disc model and Gaussian component fit the emission lines well, but how much of a contribution the disc component provides at later times is uncertain, and it is possible it completely disappears. \citet{hung20} also fit an elliptical model with no Gaussian component and, despite our model being for a circular disc, their best fit parameters all agree with ours within 1-2$\sigma$, suggesting a circular disc is a reasonable approximation.

\section{Discussion} \label{discussion}
\subsection{Balmer line emission}
\begin{figure}
\includegraphics[width=\columnwidth,trim={1cm 0.2cm 1cm 1cm},clip]{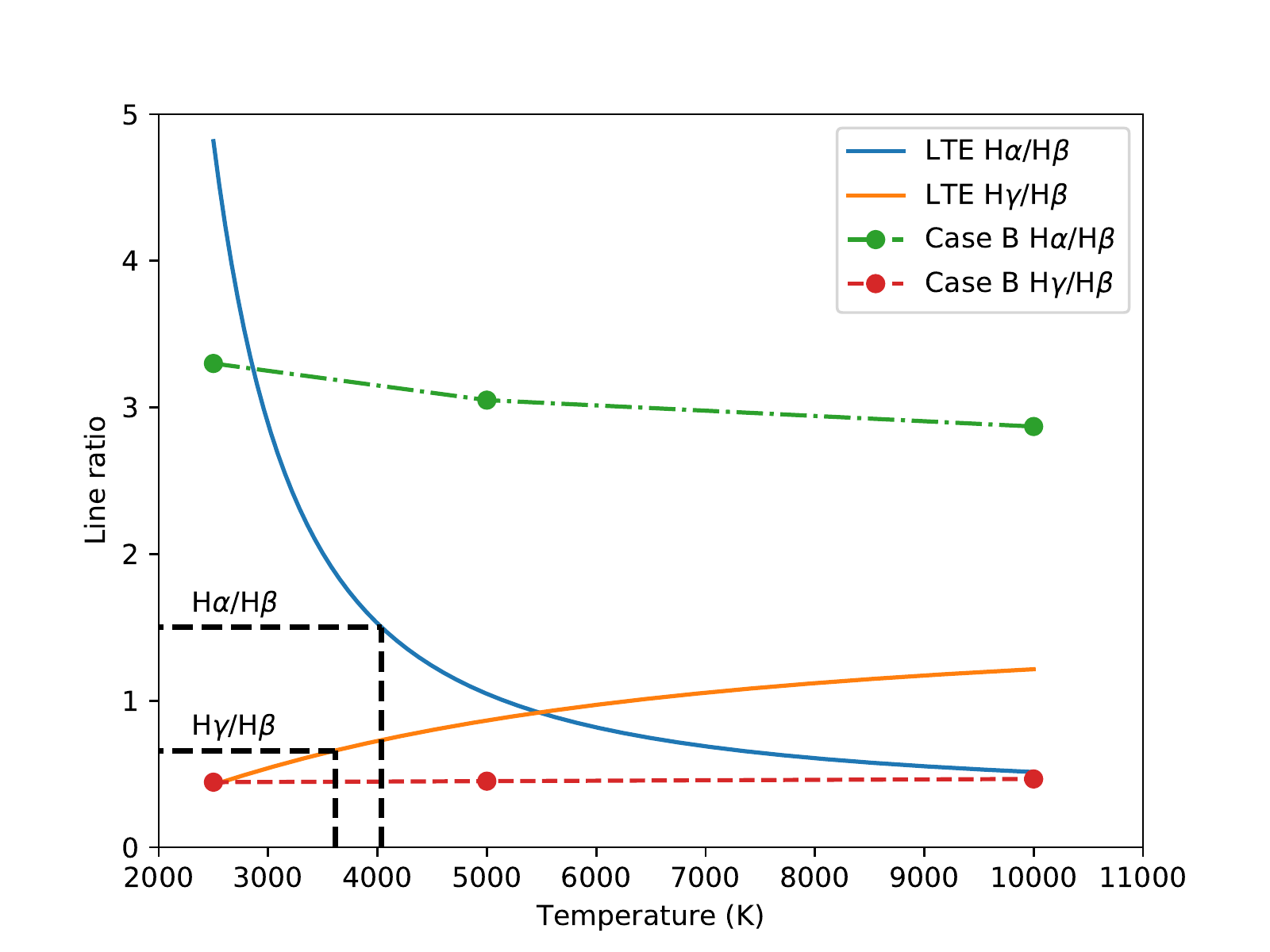}
\caption{The LTE curve was generated by dividing H$\alpha$/H$\beta$ and H$\gamma$/H$\beta$ flux ratios expected from a black-body at a given temperature. The Case B values are adopted from \protect\cite{ferost}. The black dashed lines indicate at what temperatures our measured H$\alpha$/H$\beta$ and H$\gamma$/H$\beta$ ratios intersect the corresponding LTE curve.}
\label{fig:lte}
\end{figure}

Possibly the most striking result from our observations is the flat Balmer decrement. In a typical AGN BLR the H$\alpha$/H$\beta$ ratio is $\sim3-4$ which is consistent with Case B recombination \citep{ferost} combined with reddening, high densities, high column densities and an extended ionised continuum \citep{Netz}. Figure \ref{fig:lte} compares ratios expected from Case B to those expected from local thermodynamic equilibrium (LTE). The LTE temperature curves are generated by determining the flux given by a black-body curve at the position of each emission line, and dividing for the required ratio. The ratios shown by AT\,2018hyz are completely inconsistent with Case B but agree with LTE at $\sim4000$K, suggesting these lines are due to collisional excitation rather than photoionisation. The caveat to this temperature measurement is that we haven't considered radiative transfer effects, particularly with regards to line saturation \citep{horne86}. This temperature may therefore be a lower limit.

Low Balmer decrements like this are observed in cataclysmic variables (CVs) and are attributed to thermal emission from an accretion disc chromosphere \citep{cv1}. In this scenario the bulk of the disc is optically thick (thus it looks like a black-body), while the chromosphere is optically thin in the continuum but thick in the lines. So we see lines from the chromosphere with fluxes corresponding to black-body emission, superimposed on the black-body continuum from the disc. As a disc will have both radial and vertical temperature gradients, these black-bodies are not necessarily at the same temperature.

We can test whether the temperature measured from the Balmer line ratio is consistent with the continuum temperature expected in a disc at that radius. In a simple accretion disc model, the temperature profile of the disc is described by a $T \propto R^{-3/4}$ relation: 

\begin{align}
    \centering
    T^4 = \left( \frac{3GM\dot{m}}{8\pi R^3\sigma} \right) = 1.529\times10^{22} \left(\frac{L}{L_{\rm Edd}}\right) \left(\frac{M}{10^9M_{\odot}}\right)^{-1} \left(\frac{R}{R_S}\right)^{-3} \label{eq:T}
\end{align}

Using a black hole mass of $M=10^{6}M_{\odot}$ and an Eddington ratio of 1.5, then at the initial radius of the Balmer emission, 600$R_{\rm S}$, equation \ref{eq:T} gives a temperature of $T=24000$\,K, whereas at 1800$R_{\rm S}$ (from the narrower lines at later times) we expect $T=11000$\,K. The temperature we measure from the line ratio is significantly cooler than this, at $\sim4000$\,K, but the initial temperature of 24\,000K is consistent with the black-body continuum temperature measured by \citet{gomez}. This suggests our observations can't be explained by a single temperature producing both the continuum and lines, which could be explained by a vertical temperature profile.

CV disc models can reproduce flat Balmer decrements but fail to produce He II lines \citep{cv1}, suggesting He II lines originate from a different source. It could be that the He II observed in AT 2018hyz originates from another source like debris stream collisions, although then it is not obvious why the line only appears in later spectra. Unlike He II, we observe He I following the same profile evolution as the Balmer lines, indicating that He I and He II were formed in different parts of the debris.

\begin{figure*}
\centering
\includegraphics[width=1\linewidth,trim={2cm 0cm 2cm 0cm},clip]{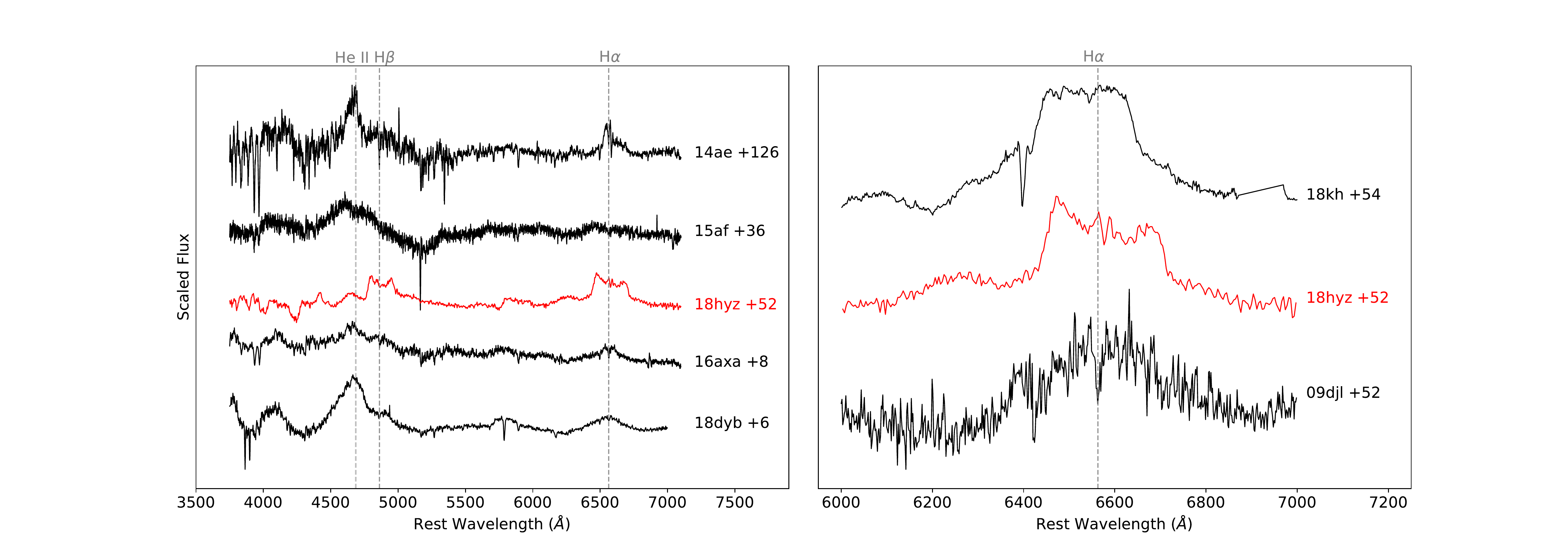}
\caption{TDE candidates with similar spectral features to AT\,2018hyz. Left: TDEs with flat Balmer decrements. Also note the late appearing He II in 14ae. Right: TDEs which may show disc-like profiles. 18hyz has the most obvious double peak.
}
\label{fig:others}
\end{figure*}

A disc origin for the Balmer lines is further supported by the double-peaked profiles that develop from $\sim$40-80 days in both H$\alpha$, H$\beta$ and possibly H$\gamma$. The appearance of these double peaks coincides with a minimum in the black-body temperature, which subsequently rises again \citep{gomez}. A possible interpretation of the temperature evolution is that there is an initial shock which starts to cool until a disc forms and heats material up again. The blue peak is consistently stronger which could be a sign of relativistic beaming. From the velocity widths of these lines we estimated that initially they originate from a region $\sim600R_{\rm S}$ from the black hole. The tidal radius, $R_{\rm t}=R_*(M_{\rm BH}/M_*)^{1/3}$, for a 1M$_{\odot}$ star around a 10$^6$M$_{\odot}$ black hole is $\sim25R_{\rm S}$. TDE accretion discs are expected to have a radius of order $R_{\rm t}$ \citep{rees88}, whereas the width of the Balmer lines corresponds to a radius of $24R_{\rm t}$. However, recent simulations by \citet{bonlu19} have shown that discs can extend out to tens of $R_{\rm t}$, which is more consistent with our measurements. If we instead assume a 0.1M$_{\odot}$ star as found by \citet{gomez}, $R_{\rm t}\sim5R_{\rm S}$, so the Balmer line widths correspond to a radius of $120R_{\rm t}$. This is less consistent but not unreasonably so, and we are taking the line widths at the beginning of observations, possibly before the disc is fully formed. We note that another possible interpretation of the double-peaked lines is that they are produced by an expanding outflow, which can provide blue- and red-shifted components as some material moves towards us and some away. This interpretation can explain the ordering of emission line appearance, and can produce the observed temperatures and velocities (\textit{Ilya Mandel, priv. communication}). However, we disfavour this interpretation as the receding part of the outflow would be obscured, so it is not obvious how we would observe a red peak. Additional evidence for the disc interpretation comes from a flattening in the UV lightcurve at $\sim550$ days after peak \citep{gomez} which is a sign of late-time accretion \citep{vv19}.

In Figure \ref{fig:cont_subbed} it is clear that a bump develops around $-12\,000$~km~s$^{-1}$ blueward of both H$\alpha$ and H$\beta$ at +36 days, and remains until at least +80 days. After this, the bump blueward of H$\alpha$ disappears while the H$\beta$ bump reddens slightly and grows. We suggest that initially we observe a fast outflow in our line of sight, similar to that described in \citet{pj94}. This then fades away, while He II starts to appear at a similar wavelength to the outflow next to H$\beta$. Another interpretation is that this blue bump is the blue wing of disc emission with smaller radius and larger velocity than the narrower double peaks. The lack of a corresponding red wing could then be due to suppression by strong relativistic beaming. If this is the case, a FWHM of $24\,000$~km~s$^{-1}$ implies emission at $\sim300R_{\rm S}$. The appearance of the blue-shifted lines coincides with a plateau in the lightcurve \citep{gomez}, suggesting these could be related. 

The double peaked profiles that develop in the Balmer lines and possibly in He I only appear temporarily before becoming round-topped again. This is difficult to explain, especially as the constant H$\alpha$/H$\beta$ ratio suggests the conditions at the emitting region do not change despite the dramatic evolution in line profile. The changes in the Balmer line profiles happen over a time consistent with the dynamical timescale at their estimated radius, meaning the evolving profiles could be due to a change in the structure of the emitting region. It could be that it takes some time for the debris to settle into a disc, at which point we then see defined double peaks, before something happens to disrupt the disc, puffing it up and resulting in rounder line profiles. However, emission line fitting by \citet{hung20} suggests the double-peaks do not completely disappear, but instead the emission becomes dominated by a Gaussian component produced by a radiatively-driven wind.

We also observe that the double peaks become less visible around the same time that He II appears, suggesting the disruption of the disc and production of He II could be related. In CVs it is suspected that He II is produced in the shocked region that forms when material from the donor star collides with the accretion disc. A similar situation could occur in a TDE if some material initially flung far out during the disruption or by shocks in stream-stream collisions eventually falls back and hits the already formed accretion disc. \citet{bonlu19} find that a constant stream of debris falling onto the accretion disc continually disrupts it and can cause the formation of spiral density waves. The velocity of material crossing the spirals could be altered which is another possible cause of change in the observed emission line profiles, though this is inconsistent with the models fit by \citet{hung20}.

\subsection{He II line emission}
The production of He II emission requires a significant flux of EUV photons at wavelengths $\leqslant228$\AA{}. While this is not consistent with the $\sim16\,000-20\,000$K temperatures estimated from photometry, \citet{gomez} find weak X-ray emission at $L_{\rm X}\sim10^{41}\rm erg s^{-1}$ indicating that there is a source of energetic photons. This means that the observed continuum is likely produced via a reprocessing layer. Though the X-ray flux throughout observations remains roughly constant, the plateau in the lightcurve is more obvious in the UV. The plateau coincides with a rise in the black-body temperature and may be consistent with the emergence of He II. This could be a sign of a change in the unobservable EUV continuum. \citet{gomez} find that the ratio between host [O~III] and X-ray emission is consistent with a weak, pre-existing AGN, so further observations are required to determine if this emission is associated with the TDE or not.

There are a number of ways that extended envelopes could form around TDEs. \citet{roth} suggest three possible sources. One is a quasi-static envelope that develops while the accretion disc is forming, though this doesn't explain why we only see He II at late times. Another option is that while we see Balmer lines formed in a disc, the He II could come from dissipation at large radii. This would explain the difference in line profile between the Balmer and He II lines. The third possibility is an outflow. This could be consistent with the late appearance of He II and be responsible for the increase in temperature. As the outflow expands it reaches the radius at which He II is thermalised, but photoionised Balmer lines are still self-absorbed. We know from reverberation mapping that AGN BLRs are stratified, and He II comes from much closer in than the Balmer lines \citep{revmap93}. \citet{metstone16} predict that a continually expanding outflow will eventually result in `ionisation-breakout', at which point X-ray and EUV photons can escape the envelope. This has been observed to occur 200-300 days after the UV/optical peak in other events \citep{15oi,azh}. The X-ray observations in \citet{gomez} extend to 232 days post-peak but show no sign of re-brightening, and late-time observations at $\sim550$ days also yield a non-detection.

\subsection{Consistency with a unified TDE model}
The reprocessing combined with the possible outflow are consistent with the unified TDE models presented by \cite{dai} and \cite{eqx}. In this scenario an accretion disc forms rapidly within a an envelope, and in the polar regions winds can develop. The envelope around the disc can be the origin of the cooler continuum observed in our spectra and trap EUV/X-ray photons. As the polar outflow expands it reaches the radius from which He~II can be produced, as described above. The He~II originating from an outflow rather than the envelope allows the late appearance of He~II to be consistent with the result found by \citet{gomez} that the envelope is contracting. The observation of He~II and blue-shifted Balmer lines suggests we are observing the transient somewhat face-on as we must be observing the polar region. The lack of photoionised lines from the envelope would suggest it is compact, as the emission lines are self absorbed at smaller radii \citep{roth}. A more compact reprocessing layer means the viewing angle into the central engine and outflows is wider. This is consistent with the inclination implied by the disc emission line profile fits.

We suspect that the collisionally excited Balmer lines are from the disc itself. We know the radius of the continuum reprocessing layer is initially $\sim10^{15}$cm from SED fits \citep{gomez} and the widths of the Balmer lines suggest they are being emitted at $\sim10^{14}$cm, implying, as expected, that the disc is smaller than the reprocessing layer. The He~II width (assuming an N~III component in blending) places its emission region at $\sim3\times10^{15}$cm, outside of the continuum radius. This is consistent with He~II arising from an outflow as it expands beyond the reprocessing layer. \citet{gomez} find that the optical depth of the continuum envelope at peak is $\sim0.8$, but that this increases to 18 at late times. The initial low optical depth may be due to the star only being partially disrupted, and the increase could explain why the Balmer emission lines lose their double-peaks, as the emitting region gets more obscured.

\subsection{Comparison to other TDE candidates}
We searched the literature for other TDE candidates with similar characteristics to AT\,2018hyz. The spectra are compared in Figure \ref{fig:others}. PS18kh \citep{ps18kh} was a TDE that showed similar line-profile evolution, initially showing roundish profiles that became boxy and possibly double-peaked over time with He II appearing at late times \citep{ps18kh2}. It had a peak luminosity of $L = 9.8\times10^{43}$ erg s$^{-1}$ and emitted a total energy $E = 3.82\times10^{50}$ erg over the observing period, similar to values measured for AT\,2018hyz \citep{gomez}. The temperature was similar at first, ranging from 14\,000 - 22\,000\,K, but at 250 days after peak it had increased to $T\gtrsim$ 50\,000K \citep{vvztf}. Also, unlike AT\,2018hyz the Balmer ratio in PS18kh is consistent with recombination, though this could be affected by reddening. \cite{ps18kh} fit elliptical disc models to the H$\alpha$ line profiles of PS18kh, finding a good fit to an evolving disc. However, \cite{ps18kh2} find good fits to a spherical outflow model, arguing the double-peaked shape is due to residuals from the host galaxy subtraction. In the case of AT 2018hyz the double peaks are much more obvious and robust against host removal.

PTF09djl showed double-peaked H$\alpha$ emission with one peak centred on the rest wavelength and the second redshfited by $3.4\times10^4$~km~s$^{-1}$ \citep{arcavi14}. disc models fit to these lines by \citet{arcavi14} required a bulk motion of $1.5\times10^4$~km~s$^{-1}$ to account for the redshifted second peak, although \cite{09djl} successfully fit a disc model with a high eccentricity at large inclination.

ASASSN-14ae \citep{14ae} is similar in that it initially only showed broad Balmer and He I emission lines with He II appearing at a later epoch. We measured the Balmer ratio of an ASASSN-14ae spectrum taken 37 days after discovery applying the same method we have used with our AT\,2018hyz spectra. We obtain a value H$\alpha$/H$\beta = 1.8\pm0.5$ which could suggest this TDE also produced collisional lines, though this was performed without host-subtraction. \citet{lelo19} found a similarly low Balmer decrement in the TDE candidate AT\,2018dyb, as well as in two other TDE candidates, iPTF15af and iPTF16axa \citep{blag19,16axa}. This suggests collisional lines may not be unusual in TDEs, and it should not be assumed that lines are always produced by photoionisation and recombination.

\section{Conclusion} \label{conclusion}
We have presented one of the best sampled spectroscopic datasets of a TDE to date (in addition to the well sampled lightcurve in \citealt{gomez}), showing AT\,2018hyz to be a strong TDE candidate with complex line evolution and revealing convincing evidence of disc formation. Measurements of velocity dispersion suggest a host black hole mass in the region of 10$^6$M$_{\odot}$, consistent with emission at $\sim L_{\rm Edd}$ at peak, similar to other TDEs.

The spectra initially show a strong, blue continuum with broad, but weak, He I and Balmer lines. Over time the blue continuum and broad lines fade but He II appears, showing evolution from an H- to He-strong TDE. The Balmer lines initially have velocity widths of 17\,000~km~s$^{-1}$ which narrow to 10\,000~km~s$^{-1}$, corresponding to Keplerian motion in regions 600$R_{\rm S}$ to 1800$R_{\rm S}$ from the black hole, respectively. The He II line has a width of 7700~km~s$^{-1}$ which places its emission at 3000$R_{\rm S}$. However, including a contribution from N III $\lambda4640$, a line produced by Bowen fluorescence, would mean He II is narrower and actually comes from a region at 10\,000$R_{\rm S}$. In either case, He II comes from a larger radius than the Balmer emission, consistent with its later appearance if the line emitting region forms dynamically.

We measure a constant ratio H$\alpha$/H$\beta$ $\sim1.5$ during the period of observations. Such a low ratio suggests emission from a region in LTE which is optically thin to continuum but thick to lines. Such conditions are seen in disc chromospheres in CVs. The resemblance to CVs is strengthened by the fact that He II seems to be produced in a different region to the Balmer lines. 

The double peaked profiles that develop in the Balmer lines and possibly He I are further evidence that the emission originates from an accretion disc around the black hole. The fact that the double peaks only appear temporarily before fading away again is difficult to explain. The timescale over which changes in the line profiles occur is consistent with the dynamical timescale at the radius associated with the Balmer lines, so the observed line evolution could be due to structural changes in the disc. However, the observed increase in optical depth suggests that obscuration may be the cause of the changing profiles. The initial low optical depth of the continuum envelope due to the possible partial disruption may be the reason we are able to see through to these double-peaked lines at all. Emission line fitting by \citet{hung20} suggests the double peaks don't completely fade away, but emission becomes wind-dominated at alter times. This disfavours a change in the disc structure being the mechanism behind the line profile evolution.

The late emerging He II could be explained by an outflow extending beyond the continuum emitting region that only reaches the radius required to produce He II at late times. This combined with the blue-shifted Balmer lines and X-ray detections are consistent with TDE unification models in which many TDEs may produce similar conditions, but what is observed depends on viewing angle. In the case of AT\,2018hyz we must have a view close enough to the polar region to see some outflow and X-rays, but close enough to edge-on to still see double-peaked lines. This rather specific sight line might explain why we haven't seen quite the same spectral evolution in any other TDEs.

AT\,2018hyz exhibits perhaps the clearest double-peaked emission lines that have been observed in a TDE and provides strong observational evidence that accretion discs form in at least some TDEs and are a significant source of the observed luminosity. However, disc profiles were only visible in the emission lines for a short period during the evolution of AT 2018hyz, showing the need for high cadence spectroscopic follow-up of TDEs. As we move into the era of large TDE samples with current (ZTF; \citealt{vvztf}) and future surveys (LSST), dedicated spectroscopic observations will be crucial to unveiling the physical processes occurring in these events.

\section*{Acknowledgements}
We thank I. Mandel for insightful comments and suggestions. We thank Y. Beletsky for carrying out the Magellan observations. PS is supported by a Science \& Technology Facilities Council (STFC) studentship. MN is supported by a Royal Astronomical Society Research Fellowship. GL was supported by a research grant (19054) from VILLUM FONDEN. IA is a CIFAR Azrieli Global Scholar in the Gravity and the Extreme Universe Program and acknowledges support from that program, from the Israel Science Foundation (grant numbers 2108/18 and 2752/19), from the United States - Israel Binational Science Foundation (BSF), and from the Israeli Council for Higher Education Alon Fellowship. TW is funded in part by European Research Council grant 320360 and by European Commission grant 730980. JB and DH are supported by NASA grant 80NSSC18K0577. NCS acknowledges support by the Science and Technology Facilities Council (STFC), and from STFC grant ST/M001326/1. KH acknowledges support from STFC grant ST/R000824/1. MG is supported by the Polish NCN MAESTRO grant 2014/14/A/ST9/00121. NI was partially supported by Polish NCN DAINA grant No. 2017/27/L/ST9/03221. LG was funded by the European Union's Horizon 2020 research and innovation programme under the Marie Sk\l{}odowska-Curie grant agreement No. 839090. KM acknowledges support from ERC Starting Grant grant no. 758638. FO acknowledge the support of the H2020 European HEMERA program, grant agreement No 730970. TEMB was funded by the CONICYTPFCHA/DOCTORADOBECAS CHILE/2017-72180113. 
This work is based on observations collected at the European Organisation for Astronomical Research in the Southern Hemisphere under ESO programmes 1103.D-0328 and 0104.B-0709 and as part of ePESSTO (the  Public ESO Spectroscopic Survey for Transient Objects Survey); observations from the Las Cumbres network; data gathered with the 6.5 meter Magellan Telescopes located at Las Campanas Observatory, Chile; observations obtained at the MDM Observatory, operated by Dartmouth College, Columbia University, Ohio State University, Ohio University, and the University of Michigan; service-mode observations (proposal ID: SW2019/P7) made with the WHT operated on the island of La Palma by the Isaac Newton Group of Telescopes in the Spanish Observatorio del Roque de los Muchachos of the Instituto de Astrof\'isica de Canarias. This research has made use of the NASA/ IPAC Infrared Science Archive, which is operated by the Jet Propulsion Laboratory, California Institute of Technology, under contract with the National Aeronautics and Space Administration.

\section*{Data Availability}
All data will be made public via \href{https://wiserep.weizmann.ac.il/}{WISeREP} and the \href{https://tde.space/}{Open TDE Catalog} immediately upon acceptance.




\bibliographystyle{mnras}
\bibliography{diablo} 

\appendix
\newpage
\onecolumn
\section{Table of Observations}

\fontsize{8}{10}
\captionsetup{width=17cm}
\begin{center}
\begin{longtable}{ c c c c c c c c }
\hline
\hline
UT Date    &  MJD    &  Phase  &  Exposure Time  &  Airmass  &  Telescope + Instrument  &  FWHM  &  Slit Width\\[2pt]
 & & (Days from peak) & (s) & (Start of obs) & & (\AA{}) & (")\\[3pt]
\hline

2018 Nov 12 & 58434 & +5 & 2700 & 1.73 & OGG 2m + FLOYDS & 12 & 2.00\\[5pt]

2018 Nov 13 & 58435 & +6 & 2100 & 1.23 & Tillinghast + FAST & 3.8 & 3.00\\[5pt]

2018 Nov 15 & 58437 & +8 & 1500 & 1.64 & NTT + EFOSC2 & 16 & 1.00\\[5pt]

2018 Nov 23 & 58445 & +15 & 2700 & 1.50 & OGG 2m + FLOYDS & 12 & 2.00\\[5pt]

2018 Dec 02 & 58454 & +24 & 1500 & 1.71 & NTT + EFOSC2 & 16 & 1.00\\[5pt]

2018 Dec 03 & 58455 & +25 & 2700 & 1.17 & OGG 2m + FLOYDS & 12 & 2.00\\[5pt]

2018 Dec 09 & 58461 & +31 & 7200 & 1.20 & MDM + OSMOS & 6.0 & 1.00\\[5pt]

2018 Dec 10 & 58462 & +32 & 2700 & 1.07 & OGG 2m + FLOYDS & 12 & 2.00\\[5pt]

2018 Dec 11 & 58463 & +33 & 1800 & 1.15 & Tillinghast + FAST & 3.8 & 3.00\\[5pt]

2018 Dec 15 & 58467 & +36 & 1500 & 1.78 & NTT + EFOSC2 & 16 & 1.00\\[5pt]

2018 Dec 16 & 58468 & +37 & 2700 & 1.07 & OGG 2m + FLOYDS & 12 & 2.00\\[5pt]

2018 Dec 22 & 58474 & +43 & 2700 & 1.08 & OGG 2m + FLOYDS & 12 & 2.00\\[5pt]

2018 Dec 28 & 58480 & +49 & 2700 & 1.77 & COJ 2m + FLOYDS & 12 & 2.00\\[5pt]

2018 Dec 31 & 58483 & +52 & 1800 & 1.17 & NTT + EFOSC2 & 16 & 1.00\\[5pt]

2019 Jan 05 & 58488 & +56 & 2700 & 1.23 & OGG 2m + FLOYDS & 12 & 2.00\\[5pt]

2019 Jan 10 & 58493 & +61 & 3600 & 1.16 & OGG 2m + FLOYDS & 12 & 2.00\\[5pt]

2019 Jan 12 & 58495 & +63 & 1500 & 1.17 & NTT + EFOSC2 & 16 & 1.00\\[5pt]

2019 Jan 21 & 58504 & +72 & 3600 & 1.26 & OGG 2m + FLOYDS & 12 & 2.00\\[5pt]

2019 Jan 24 & 58507 & +75 & 1500 & 1.17 & NTT + EFOSC2 & 16 & 1.00\\[5pt]

2019 Jan 27 & 58510 & +77 & 7200 & 1.22 & MDM + OSMOS & 6.0 & 1.00\\[5pt]

2019 Jan 30 & 58513 & +80 & 1800 & 1.22 & Clay + LDSS3c & 4.3 & 1.00\\[5pt]

2019 Feb 10 & 58524 & +91 & 1500 & 1.30 & NTT + EFOSC2 & 16 & 1.00\\[5pt]

2019 Feb 24 & 58538 & +104 & 1800 & 1.38 & NTT + EFOSC2 & 16 & 1.00\\[5pt]

2019 Mar 05 & 58547 & +113 & 3600 & 1.52 & OGG 2m + FLOYDS & 12 & 2.00\\[5pt]

2019 Mar 09 & 58551 & +117 & 7200 & 1.20 & MDM + OSMOS & 6.0 & 1.00\\[5pt]

2019 Mar 16 & 58558 & +123 & 3600 & 1.15 & OGG 2m + FLOYDS & 12 & 2.00\\[5pt]

2019 Mar 16 & 58558 & +123 & $2\times 1800$ & 1.20 + 1.17 & NTT + EFOSC2 & 16 & 1.00\\[5pt]

2019 Mar 23 & 58565 & +130 & 920 & 1.18 & VLT + X-SHOOTER & 0.88 & 0.90 \\[5pt]

2019 Mar 29 & 58571 & +136 & 3600 & 1.07 & OGG 2m + FLOYDS & 12 & 2.00\\[5pt]

2019 Mar 31 & 58573 & +138 & 1800 & 1.08 & OGG 2m + FLOYDS & 12 & 2.00\\[5pt]

2019 Apr 07 & 58580 & +144 & 3600 & 1.66 & Baade + IMACS & 4.0 & 1.00\\[5pt]

2019 Apr 09 & 58582 & +146 & 3600 & 1.11 & OGG 2m + FLOYDS & 12 & 2.00\\[5pt]

2019 Apr 12 & 58585 & +149 & 900 & 1.16 & IdP + B\&C & 2.0 & 1.65\\[5pt]

2019 Apr 20 & 58593 & +157 & 3600 & 1.08 & OGG 2m + FLOYDS & 12 & 2.00\\[5pt]

2019 Apr 25 & 58598 & +162 & $2\times2700$ & 1.50 + 1.89 & NTT + EFOSC2 & 16 & 1.00\\[5pt]

2019 May 10 & 58613 & +176 & 900 & 1.18 & IdP + B\&C & 2.0 & 1.65\\[5pt]

2019 May 13 & 58616 & +179 & 2700 & 1.76 & NTT + EFOSC2 & 16 & 1.00\\[5pt]

2019 Jun 07 & 58641 & +203 & 2400 & 1.66 & Baade + IMACS & 4.0 & 1.00\\[5pt]

2019 Jun 10 & 58644 & +206 & 3600 & 1.33 & NTT + EFOSC2 & 16 & 1.50\\[5pt]

2019 Nov 28 & 58815 & +369 & $2\times2700$ & 1.80 + 1.61 & NTT + EFOSC2 & 16 & 1.00\\[5pt]

2020 Jan 03 & 58851 & +404 & $2\times2700$ & 1.17 + 1.18 & NTT + EFOSC2 & 16 & 1.00\\[5pt]

2020 Jan 06 & 58854 & +406 & 2700 & 1.12 & WHT + ISIS & 1.6 & 1.00 \\[5pt]

2020 Jan 26 & 58874 & +426 & 1300 & 1.22 & VLT + X-SHOOTER & 0.88 & 0.90 \\[5pt]

\caption{Summary of all observations used in this paper. X-Shooter setup shown is for VIS arm. For UVB arm: exp time = 1360s, FWHM = 0.80\AA{} and slit width = 1.0". For NIR arm: exp time = 13x100s, FWHM = 3.1\AA{} and slit width = 0.9".}
\label{tab:obs}
\end{longtable}
\end{center}


\bsp	
\label{lastpage}
\end{document}